\documentclass[review]{elsarticle}
\usepackage{mdframed}
\usepackage{amsmath,amssymb}
\usepackage{lipsum}
\usepackage{bm}
\usepackage{graphicx}
\usepackage{graphics}
\usepackage{amsmath}
\usepackage{array}
\usepackage{graphicx}
\usepackage[caption=false,farskip=0pt,labelfont={bf}]{subfig}
\usepackage{xcolor}
\usepackage{subfig}
\usepackage{hyperref}
\usepackage[capitalise]{cleveref}
\usepackage{textcomp}
\usepackage{epstopdf}
\usepackage{cases}
\usepackage{amssymb}
\usepackage{multirow}
\usepackage{siunitx}
\usepackage{cases}
\usepackage{tabularx}
\usepackage[top=1.1in, bottom=1.1in, left=1.0in, right=1.0in]{geometry}
\usepackage{tikz}
\usepackage{comment}
\usepackage{lineno,hyperref}
\modulolinenumbers[5]
\usepackage{booktabs}
\renewcommand{\arraystretch}{1.0} 

\begin{document}

\title{Semi-implicit Lax-Wendroff kinetic scheme for multi-scale phonon transport}
\author[add1]{Shuang Peng}
\ead{15831129397@163.com}
\author[add2]{Songze Chen}
\ead{chensz@tenfong.cn}
\author[add1]{Hong Liang\corref{cor1}}
\ead{lianghongstefanie@163.com}
\author[add1]{Chuang Zhang\corref{cor1}}
\ead{zhangc520@hdu.edu.cn}
\address[add1]{Department of Physics, School of Sciences, Hangzhou Dianzi University, Hangzhou 310018, China}
\address[add2]{TenFong Technology Company, Nanshan Zhiyuan, No. 1001, Xueyuan Avenue, Taoyuan Street, Nanshan District, Shenzhen, China}
\cortext[cor1]{Corresponding author}
\date{\today}

\begin{abstract}

Fast and accurate predictions of the spatiotemporal distributions of temperature are crucial to the multi-scale thermal management and safe operation of microelectronic devices.
To realize it, an efficient semi-implicit Lax-Wendroff kinetic scheme is developed for numerically solving the transient phonon Boltzmann transport equation (BTE) from the ballistic to diffusive regime.
The biggest innovation of the present scheme is that the finite difference method is used to solve the phonon BTE for the reconstruction of the interfacial distribution function at the half-time step, where the second-order numerical schemes are used for both the temporal and spatial discretization.
Consequently, the phonon scattering and migration are coupled together within one time step, and the evolution process of phonon distribution function follows the actual physical law even if the time step is much longer than the relaxation time.
Numerical results show that the present scheme could accurately predict the steady/unsteady heat conduction in solid materials from the ballistic to diffusive regime, and its time step or cell size is not limited by the relaxation time or phonon mean free path.
The present work could provide a useful tool for the efficient predictions of the macroscopic spatiotemporal distributions in the multi-scale thermal engineering.

\end{abstract}


\maketitle

\section{Introduction}

Faster, cheaper and more accurate multi-scale thermal simulations are becoming more and more important in the industrial production, especially with the rapid development of chip semiconductor process technology, new energy vehicles, battery, aerospace and other fields~\cite{science_2022_moorelaw_ahead,TCAD_application_intel_2021_review,moore_emerging_2014,warzoha_applications_2021,pop_energy_2010,HUA2023chapter}. 
Compared with the expensive experimental inspection, the low-cost and efficient thermal simulation can predict the thermal environment in advance, and effectively mitigate the hotspot issues in the electronic devices~\cite{generation_thermalFET,warzoha_applications_2021,pop_energy_2010,rising_temperature2021,HUA2023chapter}.
However, the multi-scale, multi-variables, multi-fields, nonlinear characteristics of complex thermal systems bring huge challenges to both the physical models and numerical algorithms.
Taking the heat dissipation in chip as an example, the spatial scale spans $7$ orders of magnitude~\cite{TCAD_application_intel_2021_review,IMEC_AI_thermal_2023,intel_2023_GAAFET}, from the electron-phonon coupling~\cite{3DFINFETtransient} in the FinFET channel region (about $10$ nm) to the micro/nano scale phonon heat conduction in the silicon fin, silicon dioxide insulation later or silicon substrate region ($100$ nm- $10~\mu$m)~\cite{3DFINFET_2014_mc}, and then to the liquid cooling micro channels~\cite{van2020co} or heat pipe regions ($10~\mu$m-$100$ mm). 
Classical Fourier's law is hard to accurately capture phonon ballistic or non-diffusive transport, and atomic molecular dynamics or quantum mechanics methods are difficult to meet the needs of industrial simulations due to unaffordable computational cost~\cite{ChenG05Oxford}. 
In contrast, the phonon Boltzmann transport equation (BTE) becomes an effective model to describe the multi-scale heat conduction in the semi-conductor electronic devices~\cite{experiments_2020_transient,YangRg05BDE,TCAD_application_intel_2021_review,barry2022boltzmann,mazumder_boltzmann_2022,esee8c149,MurthyJY05Review,Chuang17gray,ZHANG20191366,zhang2021e,ZHANG2023124715}.

Unfortunately, there are no analytical solutions of BTE for most heat conduction problems, and it is not an easy task to accurately and quickly obtain the spatiotemporal distributions of temperature by numerically solving the unsteady BTE.
On one hand, the BTE is an integro-partial differential equations with multiple variables.
On the other hand, it is difficult for many methods which numerically solve the BTE to efficiently simulate heat conduction problem from the ballistic to diffusive regime due to the limitations of numerical accuracy, convergence, stability, asymptotic property and computational efficiency~\cite{ADAMS02fastiterative,Dimarco_Pareschi_2014,Jin_2022_Acta_Numerica,PhysRevE.107.025301,guo_progress_DUGKS,ZHANG2023124715,Chuang17gray,Numericalanalysis}. 
For example, in the typical direct simulation Monte Carlo statistics method~\cite{Bird_1963_POF,Bird_1978_ARFM,DSMC_book_1994}, the evolution processes of a lot of statistics particles are simulated in the phase space and the associated macroscopic fields are obtained by statistical averaging of each particle information.
G. A. Bird did lots of pioneering and outstanding work in the Monte Carlo method for rarefied gas flow  around a blunt body in the 1960s and 1970s~\cite{Bird_1963_POF,Bird_1978_ARFM,DSMC_book_1994}.
This famous method has also been progressed to solve the phonon BTE in the past $30$ years~\cite{DSMC_phonon_1994,PhysRevB.38.7576,MazumderS01MC,randrianalisoa2008monte,PJP11MC,PATHAK2021108003}.
This method can deal with complex multi-field coupling heat transfer or fluid flow problems, but it suffers from 
huge computational amount and statistical noises when the system size or scales increase sharply.
In addition, its cell size or time step is limited by mean free path or relaxation time, which makes this method difficult to be applied successfully in the (near) diffusive heat transfer or continuous fluid flow~\cite{Bird_1978_ARFM,MazumderS01MC,PJP11MC}.
To overcome this problem, the BTE-Fourier hybrid method is used~\cite{JAP_qinghao_2017,Pareekshith16BallisticDiffusive,mittal2011,MurthyJY12HybridFBTE,li2018b}, where the phase space is usually divided into several subdomains and buffer areas based on the engineers experience of engineers.
The BTE or Fourier solvers are used for different subdomain, respectively, depending on whether the heat conduction is closer to ballistic or diffusive regime.
For the buffer areas, iterative or overlapping strategies are usually used to realize the numerical coupling of the BTE and Fourier solvers.
Although this method is widely used in thermal engineering, the coupling between different numerical algorithms is not easy, and the division of geometric regions or the application range of Fourier's law is empirical or hard to draw boundaries. 
Actually different experience values and phase space divisions strategies significantly affect the final results.

Apart from above methods, some researchers focused on developing unified kinetic schemes with asymptotic/unified property~\cite{Dimarco_Pareschi_2014,Jin_2022_Acta_Numerica,PhysRevE.107.025301} to realize the efficient multi-scale simulations from ballistic to diffusive limits, for example synthetic iterative scheme~\cite{DSAnuclear,ADAMS02fastiterative,Chuang17gray,ZHANG20191366,zhang2021e,ZHANG2023124715,Hu_2024}, (discrete) unified gas kinetic scheme~\cite{XuK10UGKS,KunXu_2021,kunxu_UGKS2015,GuoZl13DUGKS,GuoZl16DUGKS,guo_progress_DUGKS}. 
On one hand, the unified kinetic scheme requires that the discretized BTE could recover the discretized macroscopic continuous/diffusion equation in the continuous/diffusive limit~\cite{Jin_2022_Acta_Numerica}. 
On the other hand, its time step or cell size should be no longer limited by relaxation time or mean free path, respectively.
The key of unified kinetic scheme is that the particle scattering and transport should be coupled together in a single time step, so that a time step much larger than relaxation time can be use in the (near) diffusive or continuous regime~\cite{KunXu_2021,PhysRevE.107.025301}.
Unified gas kinetic scheme (UGKS)~\cite{XuK10UGKS} is one such numerical method and applied successfully for multi-scale particle transport in the past $14$ years~\cite{KunXu_2021,kunxu_UGKS2015}, where both the macroscopic governing equation and BTE are introduced and solved under the framework of finite volume method.
The macroscopic flux is obtained by taking the moment of the distribution function at the cell interface instead of traditional constitutive relationship.
More importantly, the formal integral solution of BTE is introduced for the reconstruction of the interface distribution function, to realize the coupling of particle scattering and transport in a single time step, rather than direct numerical interpolation.
Considering the complex mathematical expression of the formal integral solution, a simpler strategy is used for the reconstruction of intefacial flux in the discrete unified gas kinetic scheme (DUGKS)~\cite{GuoZl13DUGKS}, namely, the BTE is solved along the characteristic line of particle transport and the midpoint rule is used for the temporal integration.
It computational efficiency is higher and mathematical expressions are easier than the UGKS, and has made great success for multi-scale heat transfer~\cite{GuoZl16DUGKS} or fluid flow problems~\cite{guo_progress_DUGKS}, too.

Except the characteristic solution or formal integral solution of BTE, actually directly using the discrete approximated solution of BTE at the cell interface is also a reasonable choice and may be easier.
Motivated by previous work of Lax-Wendroff scheme for multi-scale gas flow~\cite{pof2022_lax_wendroff_weidongli,csz_Lax_Wendroff2022}, an efficient semi-implicit Lax-Wendroff kinetic scheme is developed in this paper for numerically solving the transient phonon BTE.
Different from the DUGKS, the phonon BTE at the cell interface is solved by a finite difference scheme regardless of phonon transport direction.
Numerical results show that the present scheme could capture the transient heat conduction from the ballistic to diffusive regime and the time step is not limited by the relaxation time.

\section{Model and Numerical Algorithm}

\subsection{Phonon BTE}

Boltzmann transport equation (BTE) has played a great role in nuclear physics~\cite{DSAnuclear,ADAMS02fastiterative}, rarefied gas dynamics~\cite{Bird_1978_ARFM,DSMC_book_1994,XuK10UGKS,KunXu_2021,kunxu_UGKS2015,GuoZl13DUGKS,guo_progress_DUGKS}, chip heat dissipation~\cite{HUA2023chapter,TCAD_application_intel_2021_review} and other fields. 
Different from the classical macroscopic constitutive relation, for example, the Fourier's law, the BTE describes the space-time distributions of macroscopic physical field by capturing the evolution law of distribution function in the time, position, momentum spaces~\cite{ChenG05Oxford}.
Taking some reference variables of system, such as reference temperature, velocity, system size and specific heat, the frequency-independent dimensionless phonon BTE under the relaxation time approximation is~\cite{Chuang17gray,HUA2023chapter,ZHANG2023124715,Hu_2024} 
\begin{align}
\frac{\partial f}{\partial t}+\bm{v_g} \cdot \nabla_{\bm{x}} f =\frac{1}{\tau } \left( -f + {f}^{eq} \right)
\end{align}
where $f=f(\bm{x},\bm{v_g},t)$ is the phonon distribution function of energy density, depending on spatial position $\bm{x}$, time $t$ and group velocity $ \bm{v_g} = \left|\bm{ v_g } \right| \bm{s} $, $\bm s= \left( \cos \theta, \sin \theta \cos \varphi, \sin \theta \sin \varphi \right)$ is the unit directional vector and assumed to be isotropic for three-dimensional materials ($\theta$ is the polar angle and $\varphi$ is the azimuthal angle).
${f}^{eq}$ is the equilibrium distribution function,
\begin{align}
{f}^{eq}=\frac{CT}{4 \pi},
\label{eq:feq}
\end{align}
where $C$ is the specific heat and $T$ is the temperature.
$\tau$ is the relaxation time, namely, the average time passed between two adjacent phonon scattering. 
Similarly, the average distance passed between two adjacent phonon scattering is the phonon mean free path $\lambda=|\bm{v}_g| \tau$.
A Knudsen number is defined as the ratio between the phonon mean free path to the system characteristic length~\cite{Bird_1978_ARFM,DSMC_book_1994,ChenG05Oxford}, i.e., $\text{Kn}=\lambda / L$.
Actually the smaller the Kn is, the more frequent the phonon scattering is, which indicates that the phonons suffer a diffusive transport process.

Energy conservation is satisfied during the phonon scattering process,
\begin{align}
0= \int \frac{{f}^{eq}-f}{\tau } d\Omega ,
\end{align}
where $d\Omega$ represents the integral over the whole solid angle space.
Macroscopic variables $W$ including local energy density $U$, heat flux $\bm{q}$ and temperature $T$ can be obtained by taking the moment of distribution function,
\begin{align}
U&= \int f d\Omega , \label{eq:energy} \\
\bm{q}&= \int \bm{v_g} f d\Omega ,  \label{eq:heatflux}  \\
T &= \frac{1 }{C}  \int f d\Omega= \frac{U}{C}. \label{eq:temperature} 
\end{align}

\subsection{Semi-implicit Lax-Wendroff kinetic scheme}

\begin{figure}[htb]
\centering
\includegraphics[width=0.8\textwidth]{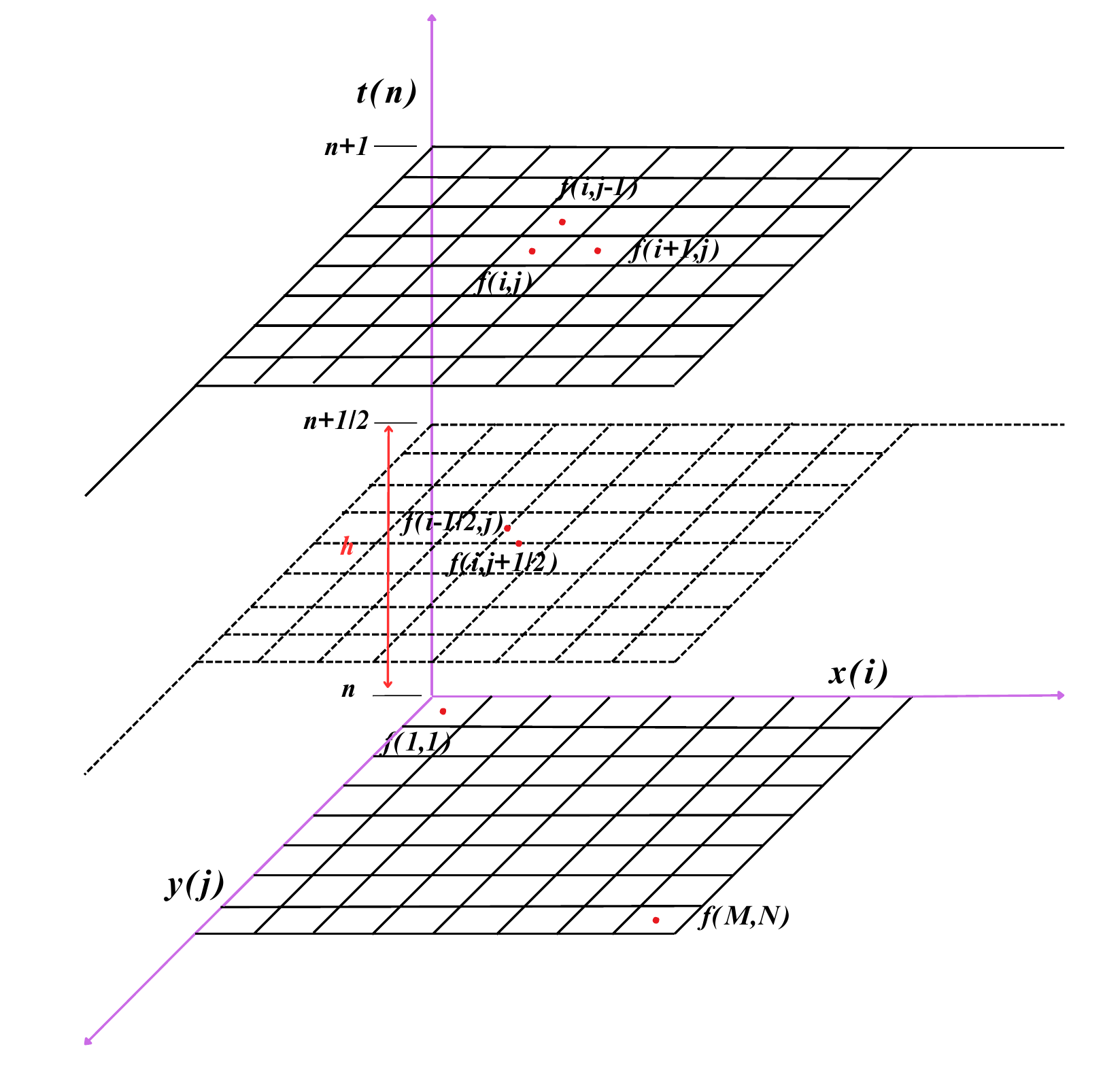}
\caption{Evolution of the phonon distribution function in the discretized spatial ($x_i,y_j$) and temporal ($t_n$) spaces from $t_n$ to $t_{n+1}=t_n +\Delta t$ for a given group velocity, where $t_{n+1/2}=t_{n}+h$, $h=\Delta t/2$. Each quadrilateral represents the discrete control volume.}
\label{evolution_xt}
\end{figure}

The semi-implicit Lax-Wendroff scheme for solving the transient phonon BTE is introduced in detail.
For ease of understanding and keeping generality, we take quasi-2D simulation as an example,
\begin{align}
\frac{\partial f}{\partial t}+u\frac{\partial f}{\partial x}+v\frac{\partial f}{\partial y} =\frac{1}{\tau } ( {f}^{eq}-f)
\end{align}
where $u=|\bm{v_g}| \cos \theta$ and $v=|\bm{v}_g| \sin \theta \cos \varphi $ are the group velocity in $x-$ and $y-$ direction, respectively.
A uniform Cartesian grid is used to discrete the spatial domain, and the temporal space is also discretized, as shown in~\cref{evolution_xt}, where $(\Delta x,~M,~i)$ and $(\Delta y,~N,~j)$ are the (cell size, total cell number, index of cell center) in the $x-$ and $y-$ direction, respectively.
The finite volume method is used to solve the transient phonon BTE at the cell center $(x_i,y_j)$ and a temporal integral from time $t^n$ to $t^{n+1}=t^{n}+\Delta t$ is performed,
\begin{align}
\int_{{t}^{n}}^{{t}^{n+1}}\frac{\partial {f}_{i,j,k}}{\partial t}dt+\int_{{t}^{n}}^{{t}^{n+1}}\left(u_k \frac{\partial {f}_{i,j,k}}{\partial x}+v_k \frac{\partial {f}_{i,j,k}}{\partial y}\right)dt=\int_{{t}^{n}}^{{t}^{n+1}}\frac{{f}_{i,j,k}^{eq}-{f}_{i,j,k}}{\tau }dt,
\end{align}
where ${f}_{i,j,k}^{n}$=$f({x}_{i},{y}_{j},u_{k},v_{k},{t}^{n})$ and the solid angle space is also discretized with index $k$.
In quasi-1D simulations, the Gauss–Legendre
quadrature is used to discrete $\theta \in [-1,1]$ into $N_{\theta}$ points.
In quasi-2D or 3D simulations, the Gauss–Legendre quadrature is also used to discretize the azimuthal angle $ \varphi \in [0,\pi]$ into $N_{\varphi} /2$ points due to symmetry.
In order to realize second-order temporal accuracy, the mid-point and trapezoidal rules are used for the temporal integral of the phonon convection and scattering terms, respectively, 
\begin{align}
\frac{{f}_{i,j,k}^{n+1}-{f}_{i,j,k}^{n}}{\Delta t}+u_k \frac{\partial {f}_{i,j,k}^{n+1/2}}{\partial x}+v_k \frac{\partial {f}_{i,j,k}^{n+1/2}}{\partial y}=\frac{1}{2}\left(\frac{{f}_{i,j,k}^{eq,n}-{f}_{i,j,k}^{n}}{\tau }+\frac{{f}_{i,j,k}^{eq,n+1}-{f}_{i,j,k}^{n+1}}{\tau }\right),
\label{eq:DBTEcenter}
\end{align}
where based on Gauss flux theorem the spatial divergence of distribution function is 
\begin{align}
\frac{\partial {f}_{i,j,k}^{n+1/2}}{\partial x} &=\frac{{f}_{i+1/2,j,k}^{n+1/2}-{f}_{i-1/2,j,k}^{n+1/2}}{\Delta x} , \label{eq:divergencex}  \\
\frac{\partial {f}_{i,j,k}^{n+1/2}}{\partial y} &=\frac{{f}_{i,j+1/2,k}^{n+1/2}-{f}_{i,j-1/2,k}^{n+1/2}}{\Delta y}, \label{eq:divergencey}  
\end{align}
where $(i \pm 1/2,j)$ or $(i,j \pm 1/2)$ represents the indexes of cell interfaces connected to cell center $(i,j)$ in the $x-$ and $y-$ direction, respectively, as shown in~\cref{evolution_xt}.

Combining above three equations (\ref{eq:DBTEcenter},\ref{eq:divergencex},\ref{eq:divergencey}) leads to the distribution function at the cell center at the next time step,
\begin{align}
{f}_{i,j,k}^{n+1} &=\frac{\tau }{\tau +h}{f}_{i,j,k}^{n}+\frac{h}{\tau +h}{f}_{i,j,k}^{eq,n+1}+\frac{ h}{  \tau +h  }({f}_{i,j,k}^{eq,n}-{f}_{i,j,k}^{n}) \notag\\
&-\frac{2h \tau }{\tau +h}\left(u_k\frac{{f}_{i+1/2,j,k}^{n+1/2}-{f}_{i-1/2,j,k}^{n+1/2}}{\Delta x}+v_k\frac{{f}_{i,j+1/2,k}^{n+1/2}-{f}_{i,j-1/2,k}^{n+1/2}}{\Delta y}\right) ,
\label{eq:function_center}
\end{align}
where $h=\Delta t/2$.
Taking the moment of Eq.~\eqref{eq:DBTEcenter} leads to the macroscopic governing equation at the cell center,
\begin{align}
{U}_{i,j,k}^{n+1}={U}_{i,j,k}^{n}-\Delta t \sum_{k} \left(u_k \frac{{f}_{i+1/2,j,k}^{n+1/2}-{f}_{i-1/2,j,k}^{n+1/2}}{\Delta x}+v_k \frac{{f}_{i,j+1/2,k}^{n+1/2}-{f}_{i,j-1/2,k}^{n+1/2}}{\Delta y}\right) {\phi }_{k},
\label{eq：W_next_center}
\end{align}
where $\sum_{k}$ is the numerical quadrature over the discretized solid angle space and ${\phi }_{k}$ is the associated weight.

Note that the equilibrium state $f^{eq}$ is totally determined by the macroscopic temperature $T$ or energy density $U$ based on Eq.~\eqref{eq:temperature}, so that the key to solve above two equations (\ref{eq:function_center},\ref{eq：W_next_center}) is to calculate the phonon distribution function at the cell interface at the half-time step $t^{n+1/2}$.
A temporal integral from $t^{n}$ to $t^{n+1/2}=t^{n}+h$ is implemented for the phonon BTE at the cell interface $(x_{i+1/2},y_j)$ and $(x_i,y_{j+1/2})$, respectively, as shown in~\cref{evolution_xt},
\begin{align}
\int_{{t}^{n}}^{{t}^{n+1/2}}\frac{\partial {f}_{i+1/2,j,k}}{\partial t}dt+\int_{{t}^{n}}^{{t}^{n+1/2}}\left(u_k \frac{\partial {f}_{i+1/2,j,k}}{\partial x}+v_k \frac{\partial {f}_{i+1/2,j,k}}{\partial y}\right)dt &=\int_{{t}^{n}}^{{t}^{n+1/2}}\frac{{f}_{i+1/2,j,k}^{eq}-{f}_{i+1/2,j,k}}{\tau }dt , \\
\int_{{t}^{n}}^{{t}^{n+1/2}}\frac{\partial {f}_{i,j+1/2,k}}{\partial t}dt+\int_{{t}^{n}}^{{t}^{n+1/2}}\left(u_k\frac{\partial {f}_{i,j+1/2,k}}{\partial x}+v_k\frac{\partial {f}_{i,j+1/2,k}}{\partial y}\right)dt &=\int_{{t}^{n}}^{{t}^{n+1/2}}\frac{{f}_{i,j+1/2,k}^{eq}-{f}_{i,j+1/2,k}}{\tau }dt.
\end{align}
Different from the characteristic solution or formal integral solution of BTE in the DUGKS or UGKS, a finite difference method is directly used and the backward Euler method is used to deal with the temporal integral,
\begin{align}
\frac{{f}_{i+1/2,j,k}^{n+1/2}-{f}_{i+1/2,j,k}^{n}}{h}+u_k\frac{\partial {f}_{i+1/2,j,k}^{n}}{\partial x}+v_k\frac{\partial {f}_{i+1/2,j,k}^{n}}{\partial y} &=\frac{1}{ \tau}\left({f}_{i+1/2,j,k}^{eq,n+1/2}-{f}_{i+1/2,j,k}^{n+1/2} \right), \label{eq:BTEfacex} \\
\frac{{f}_{i,j+1/2,k}^{n+1/2}-{f}_{i,j+1/2,k}^{n}}{h}+u_k\frac{\partial {f}_{i,j+1/2,k}^{n}}{\partial x}+v_k\frac{\partial {f}_{i,j+1/2,k}^{n}}{\partial y} &=\frac{1}{ \tau} \left( {f}_{i,j+1/2,k}^{eq,n+1/2}-{f}_{i,j+1/2,k}^{n+1/2} \right). \label{eq:BTEfacey}
\end{align}
Make a transformation of above two equations (\ref{eq:BTEfacex},\ref{eq:BTEfacey}) leads to
\begin{align}
{f}_{i+1/2,j,k}^{n+1/2} &=\frac{\tau }{\tau +h}{f}_{i+1/2,j,k}^{n}-\frac{\tau h}{\tau +h}\left(u_k \frac{\partial {f}_{i+1/2,j,k}^{n}}{\partial x}+v_k \frac{\partial {f}_{i+1/2,j,k}^{n}}{\partial y}\right)+\frac{h}{\tau +h}{f}_{i+1/2,j,k}^{eq,n+1/2} , \label{eq:functionface_x} \\
{f}_{i,j+1/2,k}^{n+1/2} &=\frac{\tau }{\tau +h}{f}_{i,j+1/2,k}^{n}-\frac{\tau h}{\tau +h}\left(u_k \frac{\partial {f}_{i,j+1/2,k}^{n}}{\partial x}+v_k \frac{\partial {f}_{i,j+1/2,k}^{n}}{\partial y}\right)+\frac{h}{\tau +h}{f}_{i,j+1/2,k}^{eq,n+1/2}. \label{eq:functionface_y}
\end{align}
In addition, taking the moment of Eqs.~(\ref{eq:BTEfacex},\ref{eq:BTEfacey}) leads to the macroscopic governing equation at the cell interface,
\begin{align}
{U}_{i+1/2,j}^{n+1/2} &={U}_{i+1/2,j}^{n} - h  \sum_{k}   \left(u_k\frac {\partial{f}_{i+1/2,j,k}^{n}}{\partial x}+v_k\frac{\partial {f}_{i+1/2,j,k}^{n}}{\partial y}\right){\phi }_{k} ,
\label{eq：W_half_face_x} \\
{U}_{i,j+1/2}^{n+1/2} &={U}_{i,j+1/2}^{n} - h \sum_{k}  \left( u_k\frac {\partial{f}_{i,j+1/2,k}^{n}}{\partial x}+v_k\frac {\partial{f}_{i,j+1/2,k}^{n}}{\partial y} \right) {\phi }_{k}. \label{eq：W_half_face_y}
\end{align}

In above four equations, $f^{eq,n+1/2}$ is totally determined by the energy density $U^{n+1/2}$ based on Eq.~\eqref{eq:temperature}, so that the key is to obtain the distribution function at the cell interface as well as its spatial gradients at the n-time step.
The second-order upwind scheme is used for the reconstruction of the interfacial distribution function and the details are shown as below.
When $u_k>0$,
\begin{align}
\label{eq:face_x_interpolation}
{f}_{i+1/2,j,k}^{n} &=1.5{f}_{i,j,k}^{n}-0.5{f}_{i-1,j,k}^{n} , \quad 1<i\leqslant M,  \notag \\
{f}_{3/2,j,k}^{n}&=2{f}_{1,j,k}^{n}-{f}_{1/2,j,k}^{n}.
\end{align}
When $u_k<0$, 
\begin{align}
\label{eq:face_y_interpolation}
{f}_{i-1/2,j,k}^{n} &=1.5{f}_{i,j,k}^{n}-0.5{f}_{i+1,j,k}^{n},\quad 1\leqslant i< M, \notag \\
{f}_{M-1/2,j,k}^{n} &=2{f}_{M,j,k}^{n}-{f}_{M+1/2,j,k}^{n}.
\end{align}
Note that ${f}_{1/2,j,k}^{n}$ with $u_k>0$ and ${f}_{M+1/2,j,k}^{n}$ with $u_k<0$ are obtained by boundary conditions, which will be discussed later.
For the reconstruction of interfacial distribution function in the $y$- direction. In order to get the spatial gradient of distribution function $(\partial f/\partial x, \partial f/\partial y)$ at the cell interface $(x_{i+1/2},y_j)$, the central scheme is used and the details are shown as below,
\begin{align}
\label{eq:face_gradients_interpolation}
\frac{\partial {f}_{i+1/2,j,k}^{n}}{\partial x} &=\frac{{f}_{i+1,j,k}^{n}-{f}_{i,j,k}^{n}}{\Delta x}, \quad 1\leqslant i<M,  ~  1\leqslant j\leqslant N  \notag  \\
\frac{\partial {f}_{1/2,j,k}^{n}}{\partial x}  &=\frac{{f}_{1,j,k}^{n}-{f}_{1/2,j,k}}{0.5\Delta x} , \quad 1\leqslant j\leqslant N \notag  \\
\frac{\partial {f}_{M+1/2,j,k}^{n}}{\partial x} &=\frac{{f}_{M+1/2,j,k}^{n}-{f}_{M,j,k}^{n}}{0.5\Delta x}, \quad 1\leqslant j\leqslant N  \notag  \\
\frac{\partial {f}_{i+1/2,j,k}^{n}}{\partial y}&=\frac{{f}_{i+1/2,j+1,k}^{n}-{f}_{i+1/2,j-1,k}^{n}}{2\Delta y} ,\quad 0\leqslant i\leqslant M ,~1<j<N  \notag  \\
\frac{\partial {f}_{i+1/2,1,k}^{n}}{\partial y}&=\frac{{f}_{i+1/2,2,k}^{n}-{f}_{i+1/2,1,k}^{n}}{\Delta y}, \quad  0\leqslant i\leqslant M \notag  \\
\frac{\partial {f}_{i+1/2,N,k}^{n}}{\partial y}&=\frac{{f}_{i+1/2,N,k}^{n}-{f}_{i+1/2,N-1,k}^{n}}{\Delta y}, \quad 0\leqslant i\leqslant M
\end{align}
Similar strategies can be implemented for the spatial gradient of distribution function $(\partial f/\partial x, \partial f/\partial y)$ at the cell interface $(x_i,y_{j+1/2})$. 

In summary, the procedure of the present semi-implicit Lax-Wendroff kinetic scheme can be summarized as follows
\begin{enumerate}
\item The distribution function and its spatial gradients at the cell interface at the $n$- time step are obtained by the second-order numerical interpolation scheme based on Eqs.~(\ref{eq:face_x_interpolation},\ref{eq:face_y_interpolation},\ref{eq:face_gradients_interpolation}).
\item Update sequentially the macroscopic fields and equilibrium state at the $(n+1/2)$- half-time step at the cell interface based on Eqs.~(\ref{eq：W_half_face_x},\ref{eq：W_half_face_y}) and Eqs.~(\ref{eq:temperature},\ref{eq:feq}), respectively.
\item Update the phonon distribution function at the $(n+1/2)$- half-time step at the cell interface based on Eqs.~(\ref{eq:functionface_x},\ref{eq:functionface_y}).
\item Update sequentially the macroscopic fields and equilibrium state at the $(n+1)$- next time step at the cell center based on Eq.~\eqref{eq：W_next_center} and Eqs.~(\ref{eq:temperature},\ref{eq:feq}), respectively.
\item Update the phonon distribution function at the $(n+1)$- next time step at the cell center based on Eq.~\eqref{eq:function_center}.
\end{enumerate}

It should be emphasized that the present scheme is not limited by the uniform grid or frequency-independent phonon BTE.
It can also be extended to the unstructured mesh~\cite{pof2022_lax_wendroff_weidongli}, and its momentum space can also be replaced by the phonon dispersion and scattering parameters in the whole first Brillouin zone obtained by DFT~\cite{ZHANG2023124715}, which will be done in the future.

\subsection{Boundary conditions}

Boundary conditions significantly influence the stability, accuracy and convergence of numerical simulation process. Improper boundary conditions may lead to discrepancies between simulation results and experimental data or real-world observations.
Three kinds of boundary conditions are tested in this paper, namely isothermal boundary conditions, diffusely reflecting adiabatic boundary conditions and periodic boundary conditions.
\begin{itemize}
\item[1)] Isothermal boundary condition is
\begin{align}
f({x}_{\text{wall}}, \bm{v}_g) = {f}^{eq}(\bm{x}_{\text{wall}},{T}_{\text{wall}}),\qquad \bm{v_g} \cdot \mathbf{{n}_{wall}}>0,
\end{align}
where $\bm{x}_{wall}$ represents the spatial coordinates of the boundary, and ${T}_{wall}$ represents the temperature at the boundary, $\mathbf{{n}_{wall}}$ refers to the normal unit vector that pointing from the boundary to the computational domain.
\item[2)] Diffusely reflecting adiabatic boundary requires that there is no heat flux across the boundary, and the reflected phonon distribution function are the same along each solid angle direction, i.e.,
\begin{align}
0 &=\bm{{q}_{wall}}\cdotp \mathbf{{n}_{wall}}, \notag \\
&= {\int}_{\bm{v_g}\cdotp \mathbf{{n}_{wall}}>0}\bm{v_g}\cdotp \mathbf{{n}_{wall}}  f(\bm{x}_{wall},\bm{v}_g) d\Omega +{\int}_{\bm{v_g}\cdotp \mathbf{{n}_{wall}}<0}\bm{v_g}\cdotp \mathbf{{n}_{wall}}  f(\bm{x}_{wall},\bm{v}_g) d\Omega , \\
\Longrightarrow &
 f(\bm{x}_{wall},\bm{v}_g)  = -\frac{{\int}_{\bm{v_g}\cdotp \mathbf{{n}_{wall}}<0}\bm{v_g}\cdotp \mathbf{{n}_{wall}}f({x}_{wall)}d\Omega }{{\int}_{\bm{v_g}\cdotp \mathbf{{n}_{wall}}>0}\bm{v_g}\cdotp \mathbf{{n}_{wall}}d\Omega } ,\qquad \bm{v_g} \cdot \mathbf{{n}_{wall}}>0 .
\end{align}
\item[3)] Periodic boundary condition is
\begin{align}
f(\bm{x}_{l},\bm{v}_g)-{f}^{eq}(\bm{x}_{l},{T}_{l},\bm{v}_g) &= f(\bm{x}_{r},\bm{v}_g)-{f}^{eq}(\bm{x}_{r},{T}_{r},\bm{v}_g),\qquad \bm{v}_g\cdotp \mathbf{n}_{l} >0, \\
f(\bm{x}_{r},\bm{v}_g)-{f}^{eq}(\bm{x}_{r},{T}_{r},\bm{v}_g) &= f(\bm{x}_{l},\bm{v}_g)-{f}^{eq}(\bm{x}_{l},{T}_{l},\bm{v}_g),\qquad \bm{v}_g\cdotp \mathbf{n}_{r}>0 ,
\end{align}
where $(T_l,\bm{x}_l,\mathbf{n}_l)$ and $(T_r,\bm{x}_r,\mathbf{n}_r)$ are the (temperature, position, unit normal vector) of two associated periodic boundaries, respectively.
\end{itemize}

\begin{figure}[htb]
     \centering
    \subfloat[]{\label{1Dfilm}\includegraphics[width=0.20\textwidth]{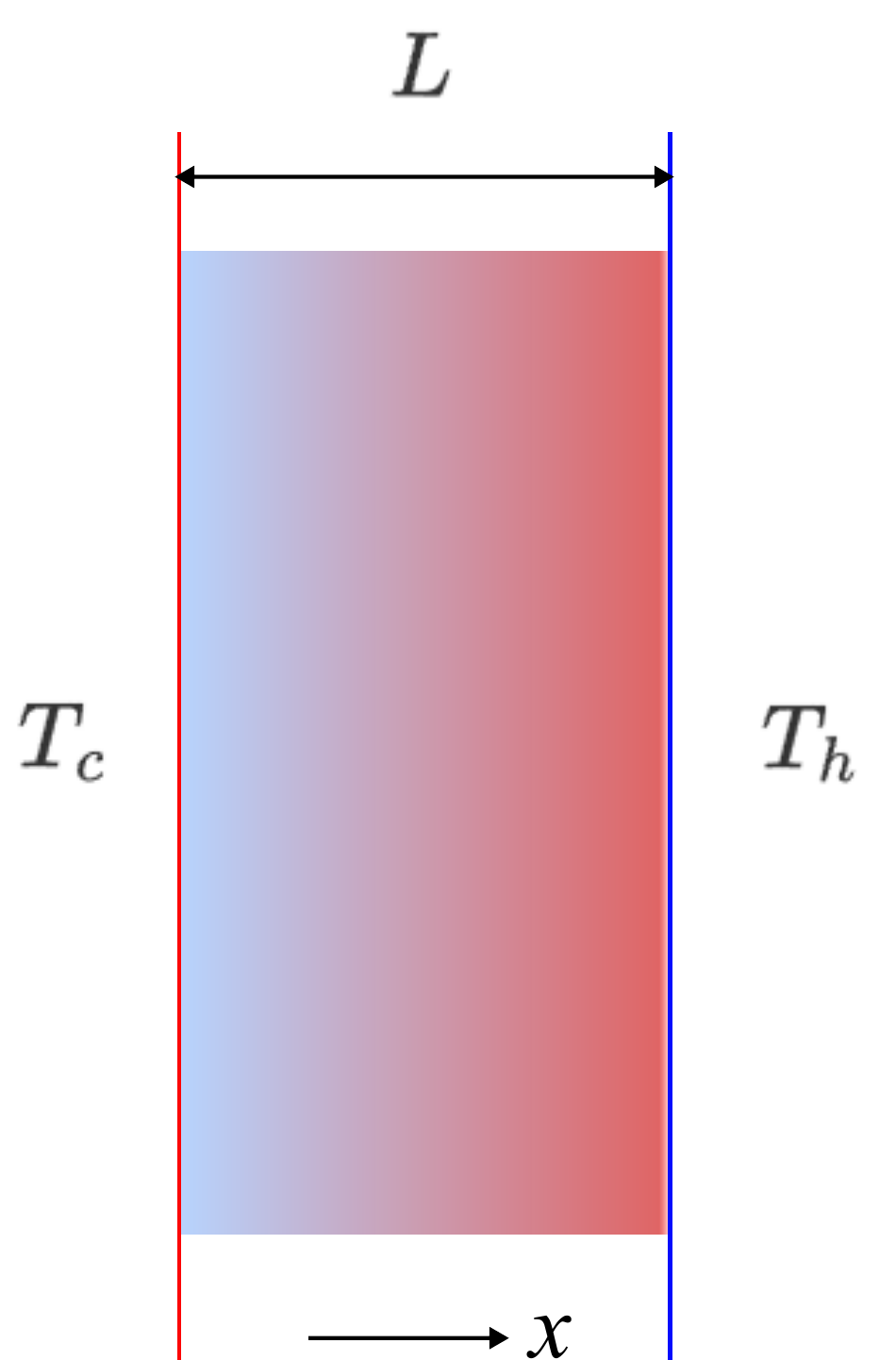}}~~~
    \subfloat[]{\label{1DTTG}\includegraphics[width=0.55\textwidth]{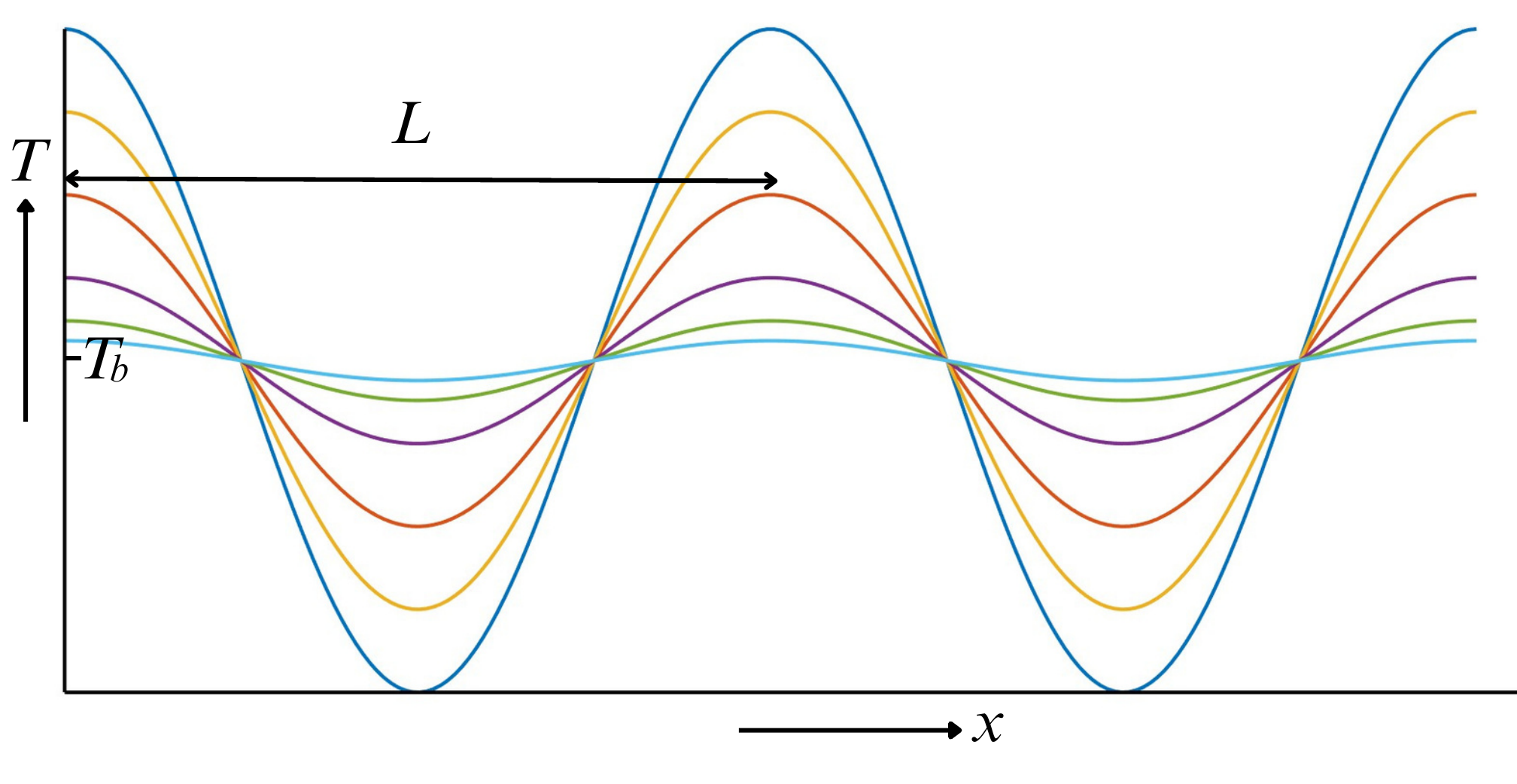}} \\
    \subfloat[]{\label{2Dinplane}\includegraphics[width=0.4\textwidth]{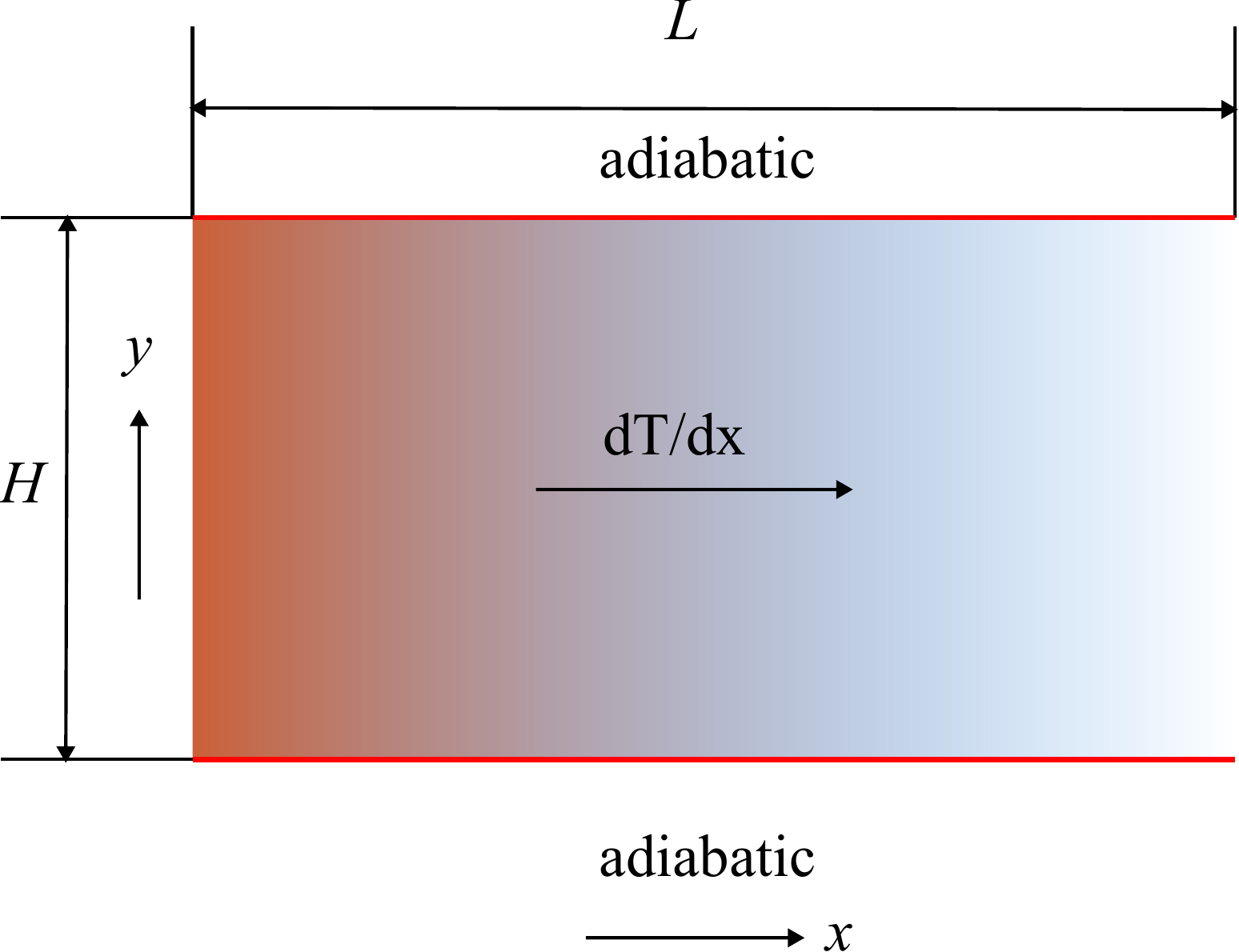}}~~~~
    \subfloat[]{\label{2Dsquare}\includegraphics[width=0.42\textwidth]{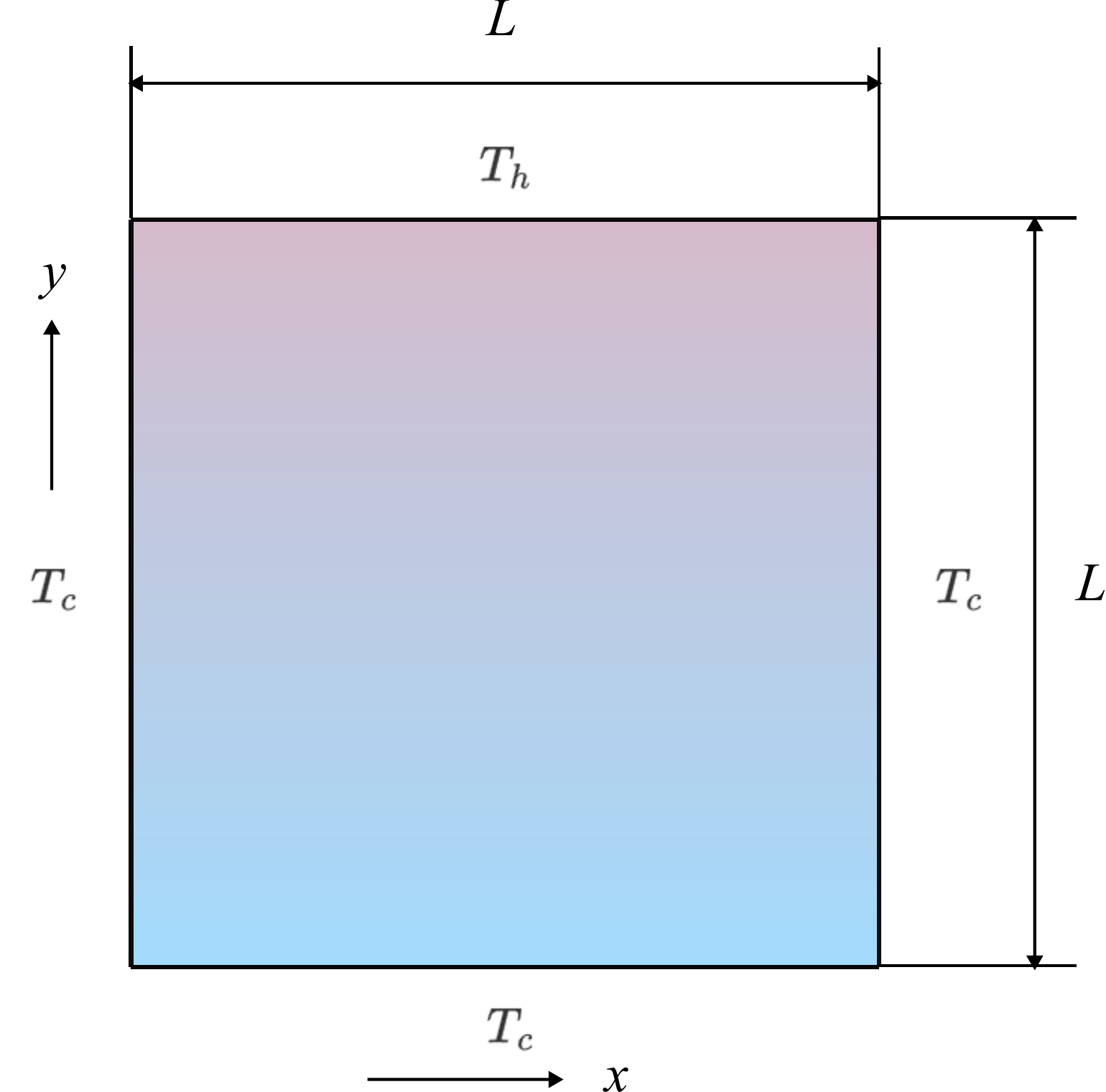}}~~\\
     \caption{Schematic of heat conduction problems. (a) Heat conduction across a quasi-1D thin film. (b) Transient thermal grating. (c) In-plane heat conduction, where the top and bottom boundaries are adiabatic. (d) Quasi-2D heat conduction in a sqaure geometry. }
     \label{four_cases_figures}
\end{figure}

\section{Numerical tests}

In order to verify the performance of semi-implicit Lax-Wendroff kinetic scheme for multi-scale phonon transport, four heat conduction problems are simulated with various Knudsen numbers, as shown in~\cref{four_cases_figures}. a) Quasi-1D heat conduction across a thin film. b) Transient thermal grating. c) Quasi-2D in-plane heat conduction. d) Quasi-2D square heat conduction.
All simulations are conducted in a personal laptop with single CPU core (13th Gen Intel(R) Core(TM) i5-13600KF 3.50 GHz).
The physical time step $\Delta t$ satisfies
\begin{align}
\Delta t = \text{CFL} \times \frac{ \{\Delta x,\Delta y \}_{\text{min}}  }{|\bm{v_g}|}
\end{align}
where $0< \text{CFL} <1$ is the Courant–Friedrichs–Lewy~\cite{Numericalanalysis,kunxu_UGKS2015} number.
Without special statements, we set dimensionless specific heat $C$ = 1, group velocity $|\bm{v_g}|$= 1, CFL=$0.40$. 

\subsection{Quasi-1D heat conduction across a thin film} 

Quasi-1D heat conduction across a thin film with thickness $L=1$ is studied, as shown in~\cref{1Dfilm}, where the temperatures at the left or right end side are ${T_h}$ and ${T_L}$, respectively. 
Isothermal boundary conditions are used for two boundaries. 
To simulate this problem, we set $N_{\theta} =24$ and $M=10$ or $M=100$ uniform cells are used to discrete the spatial domain.
The system reaches steady state when ${\epsilon }_{1} < 1.0\times {10}^{-10}$, where
\begin{align}
\epsilon _1 = \frac{1}{M}\displaystyle\sum_{i=1}^{M}\left| \frac{{W}_{i}^{n+1}-{W}_{i}^{n}}{{W}_{i}^{n}} \right|
\end{align}

\begin{figure}[htb]
     \centering
     \subfloat[]{\includegraphics[width=0.45\textwidth]{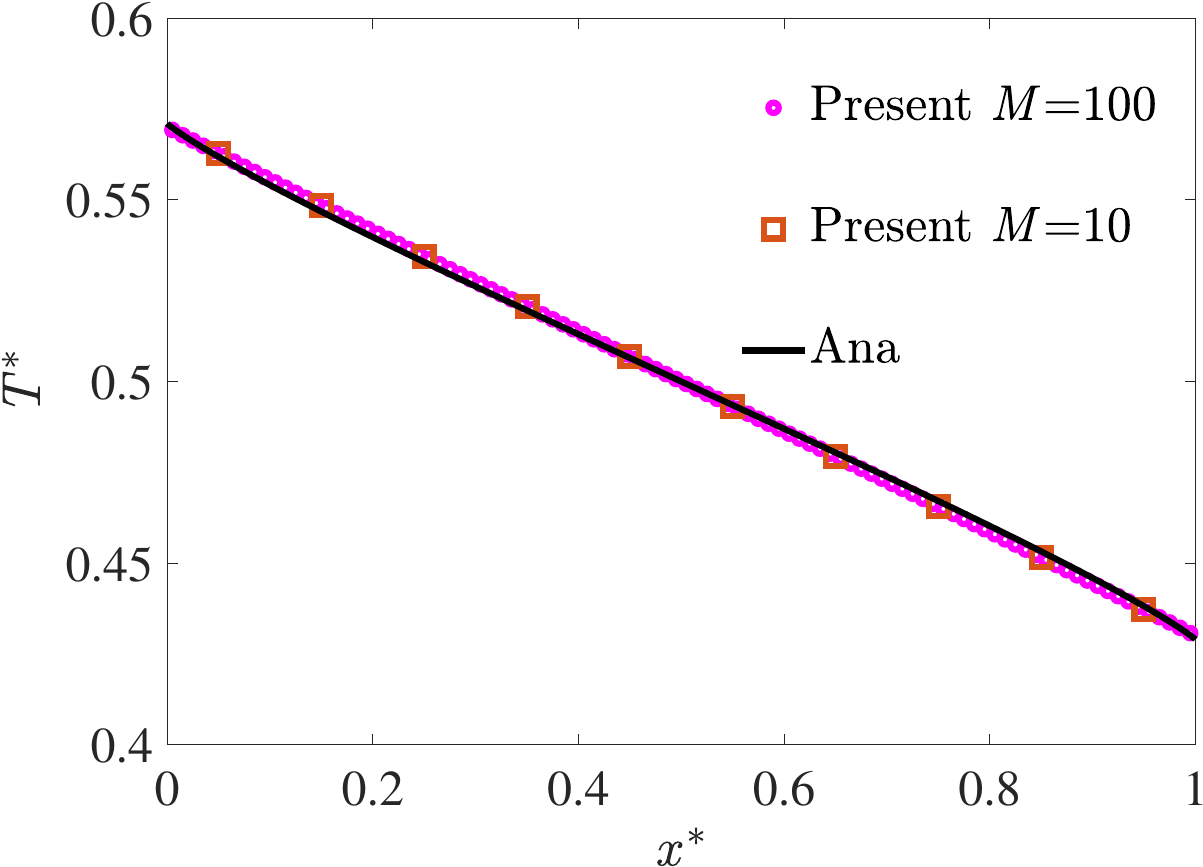}}~~
     \subfloat[]{\includegraphics[width=0.45\textwidth]{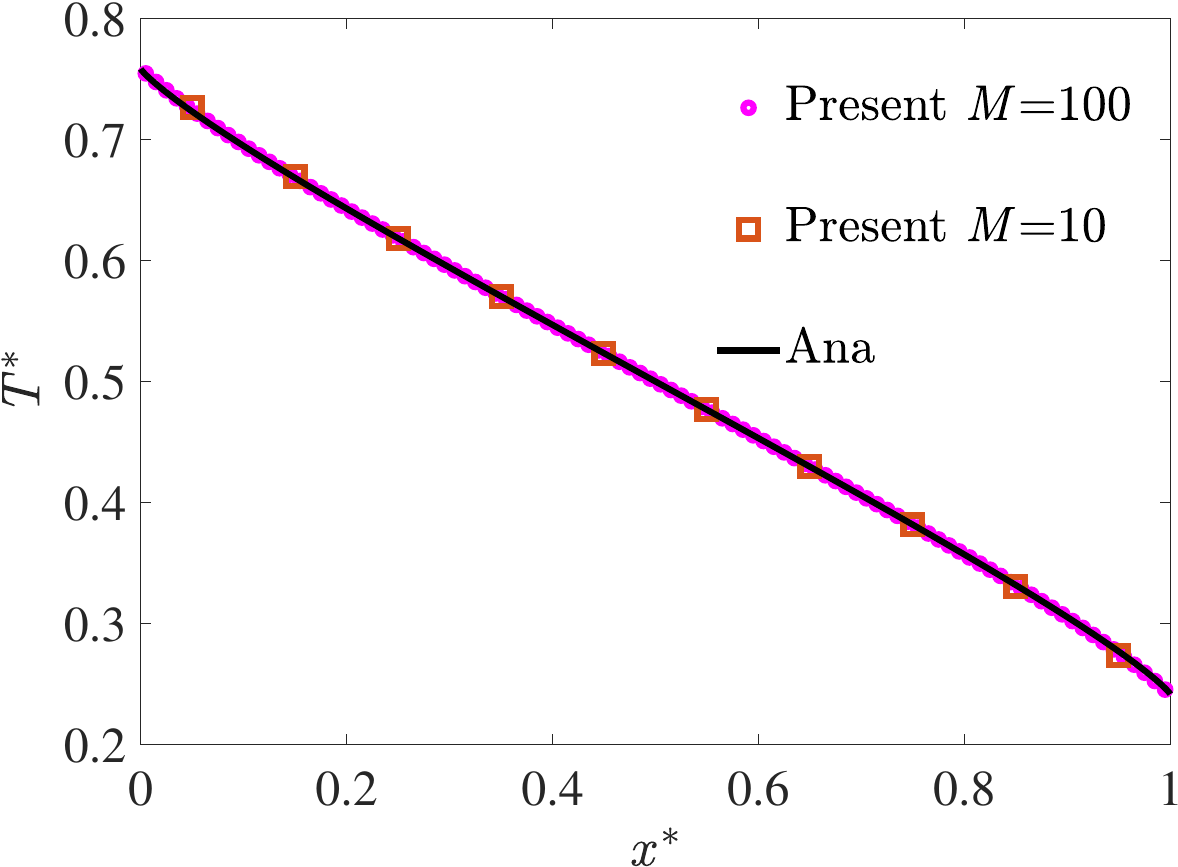}}~~\\
     \subfloat[]{\includegraphics[width=0.45\textwidth]{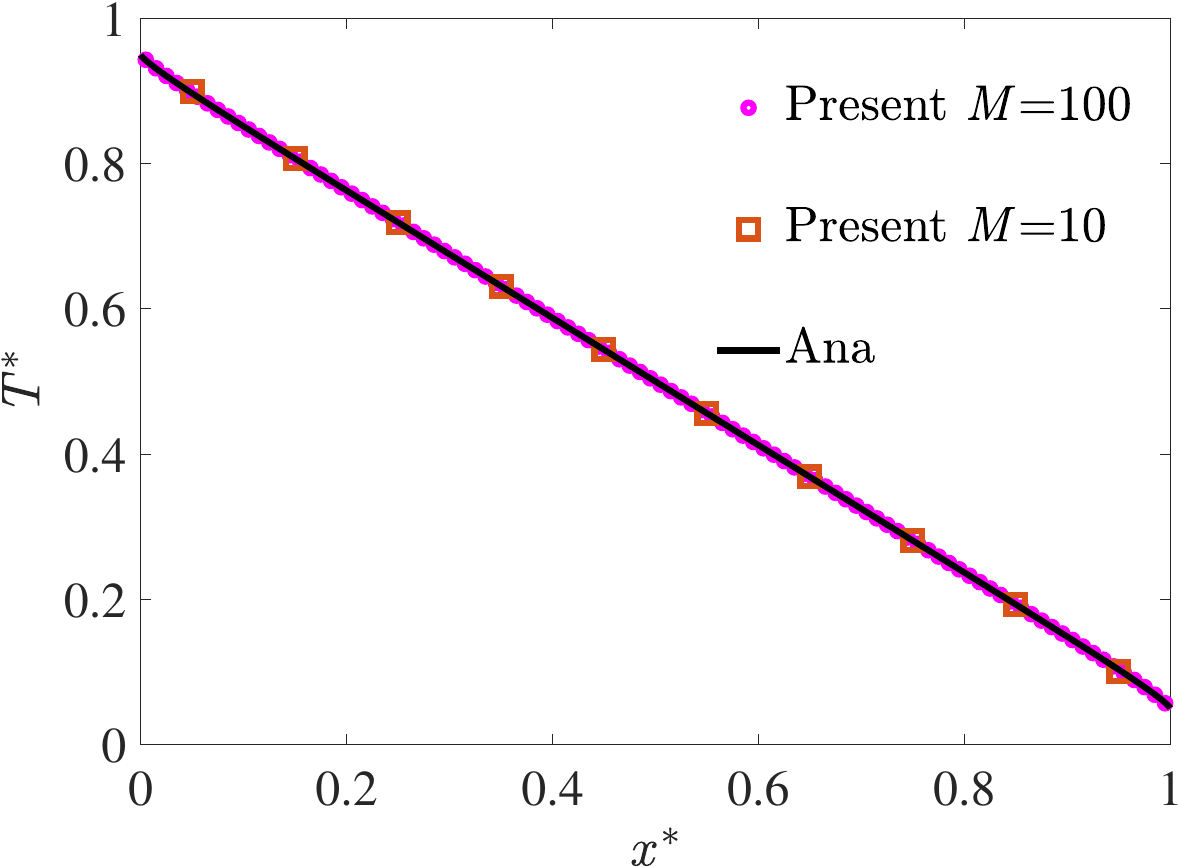}}~~
     \subfloat[]{\label{1dkn}\includegraphics[width=0.45\textwidth]{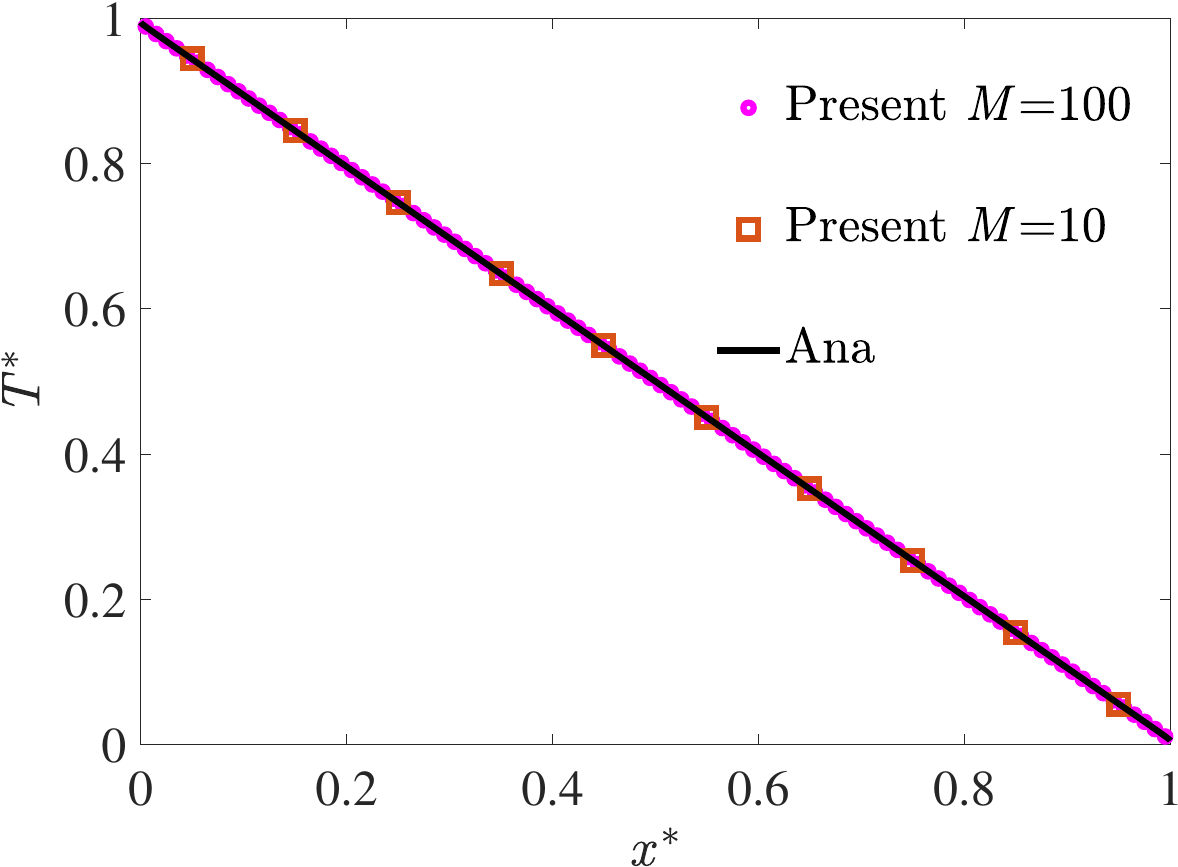}}
     \caption{Non-dimensional temperature distributions $T^* = \left( T-T_L\right)/(T_h-T_L)$ at different Knudsen numbers, where $x^*=x/L$, We use the same $\Delta t=0.04$ for different Knudsen numbers, `Ana' is the analytical solutions~\cite{MajumdarA93Film,GuoZl16DUGKS}. (a) Kn=10.0, (b) Kn=1.0, (c) Kn=0.1, (d) Kn=0.01. }
     \label{filmtemperature}
\end{figure}
The distributions of the dimensionless temperature ${T^*} = \left( T-T_L\right)/(T_h-T_L)$ at different Knudsen numbers are plotted in~\cref{filmtemperature}, where the dimensionless coordinate is $x^*=x/L$.
According to the profiles, it can be seen that the dimensionless temperature distributions predicted by the present scheme are basically in consistent with the analytical solutions~\cite{MajumdarA93Film,GuoZl16DUGKS} from the ballistic to diffusive regime.
In addition, the results with coarse grid $M=10$ are the same as those with finer grid $M=100$ for all Knudsen numbers.
When $\text{Kn}=0.01$, the present scheme can still correctly capture the heat conduction even if the cell size is much larger than the phonon mean free path ($\Delta x/\lambda=10$), or the time step is larger than the relaxation time ($\Delta t/\tau=4$).

Actually the whole spatial domain can be separated into three parts, namely, internal geometry region, Knudsen layer~\cite{kunxu_UGKS2015} and boundaries.
The thickness of the Knudsen layers is comparable to the phonon mean free path and it can be regarded as a transition layer between the internal geometry and the boundary. 
When $\text{Kn} \rightarrow 0$, the thickness of the Knudsen number is much smaller than the cell size, and its effect on the whole domain can be ignored. 
The phonon transport in the inner geometric region follows the Fourier's law and the temperature presents a linear distribution. 
Due to sufficient phonon scattering, when $\text{Kn} \rightarrow \infty$, the Knudsen layer extends over the entire geometric region. 
In this regime, the boundary scattering determines the phonon transport or temperature distribution throughout the domain. 
Phonons emitted from the isothermal boundaries hit the other boundary directly without scattering, so that the internal temperature tends to be a constant. 
When $\text{Kn} =0.1 $ or $1.0$, phonon scattering is not sufficient, and Knudsen number, boundary conditions, and internal region phonon scattering all affect the temperature distributions in the whole domain. 
From~\cref{filmtemperature}, it can be seen that the temperature in the inner geometric region is linear overall, but the temperature distribution presents a curved trend and there is temperature slip in the Knudsen layer region, which is a combined result of insufficient phonon-phonon scattering and boundary scattering. 

\subsection{Transient thermal grating}

Transient thermal grating is a sophisticated technique employed to investigate the thermal properties of materials, frequently utilized in laser heating experiments~\cite{huberman_observation_2019,PhysRevLett.110.025901}. 
This method entails the application of a short pulse laser to heat the material's surface, inducing localized temperature variations. 
Specifically, it uses crossed laser beam interference to produce a spatially cosine temperature profile in a sample (\cref{1DTTG}),
\begin{align}
T(x,0)=T_b+A_0 \cos (\alpha x),
\end{align}
where $T_b$ is the background temperature, $A_0$ is the amplitude of the temperature variation, and $\alpha =2\pi /L$ is the wave number with $L=1$ being the spatial grating period.
In this paper, we take a spatial grating period as an example to simulate the evolution process of peak temperature amplitude with time.
The periodic boundary conditions are applied at the boundary of $x=0$ and $x=L$. 

\begin{figure}[htb]
 \centering
 \subfloat[$\xi=2\pi$Kn, $\Delta t=0.005, \Delta x=0.02$]{\includegraphics[width=0.45\textwidth]{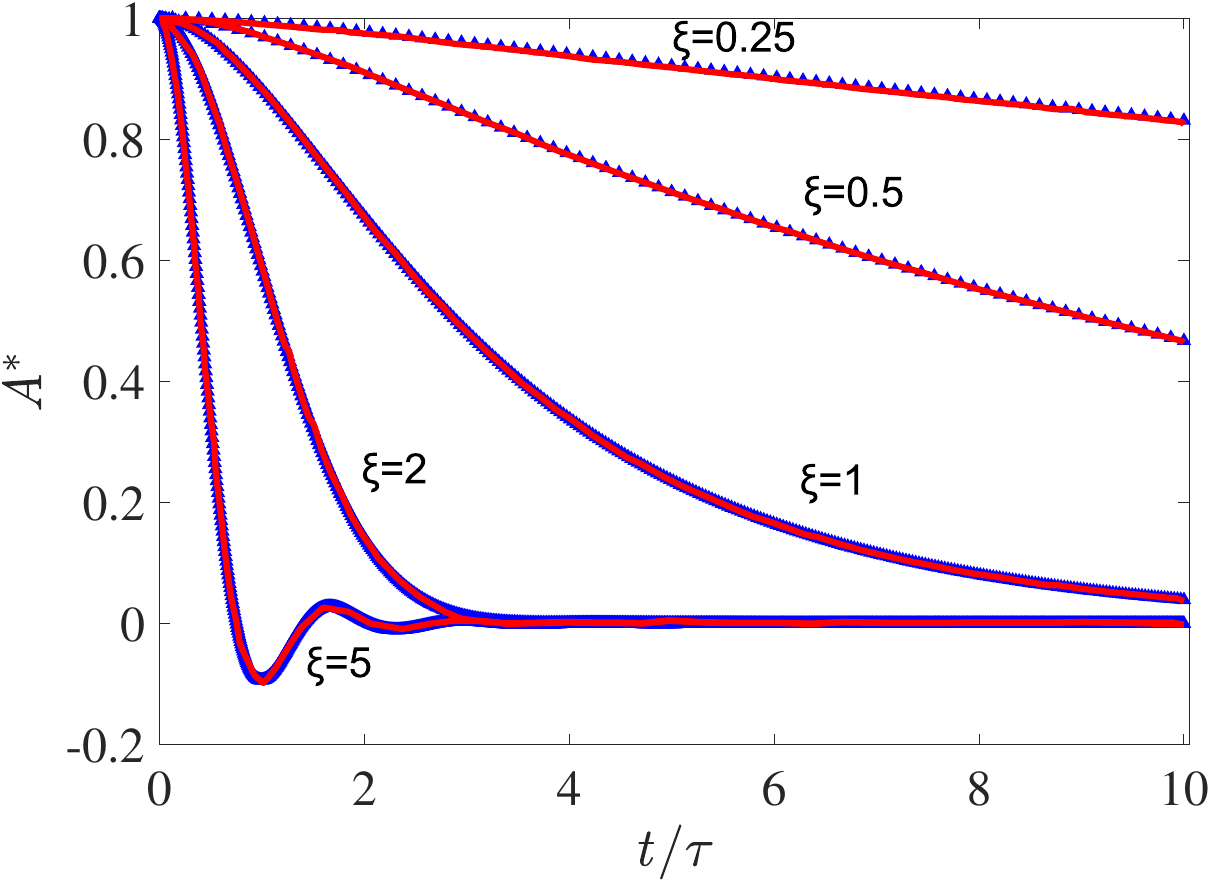}}~~
 \subfloat[Kn=0.01]{ \includegraphics[width=0.45\textwidth]{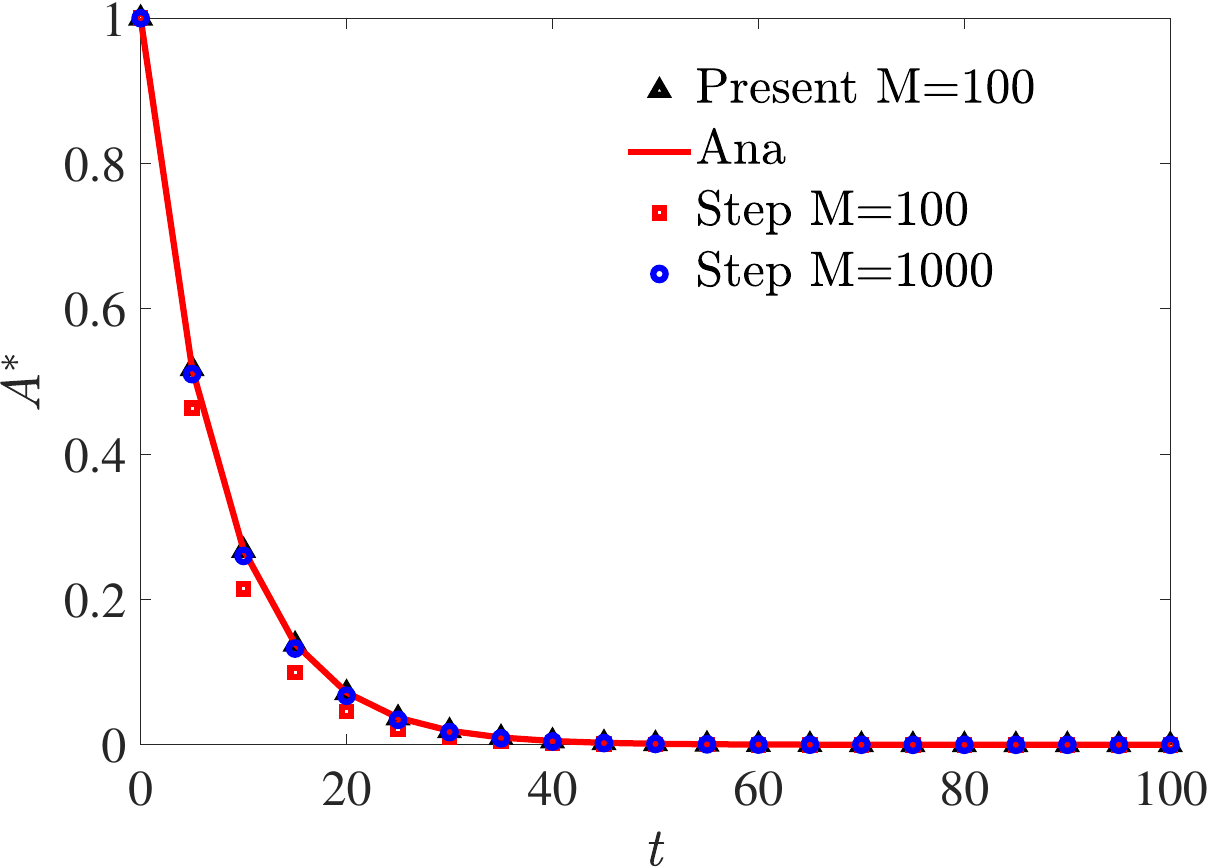} }  \\
 \subfloat[Kn=0.001]{ \label{e-3}\includegraphics[width=0.45\textwidth]{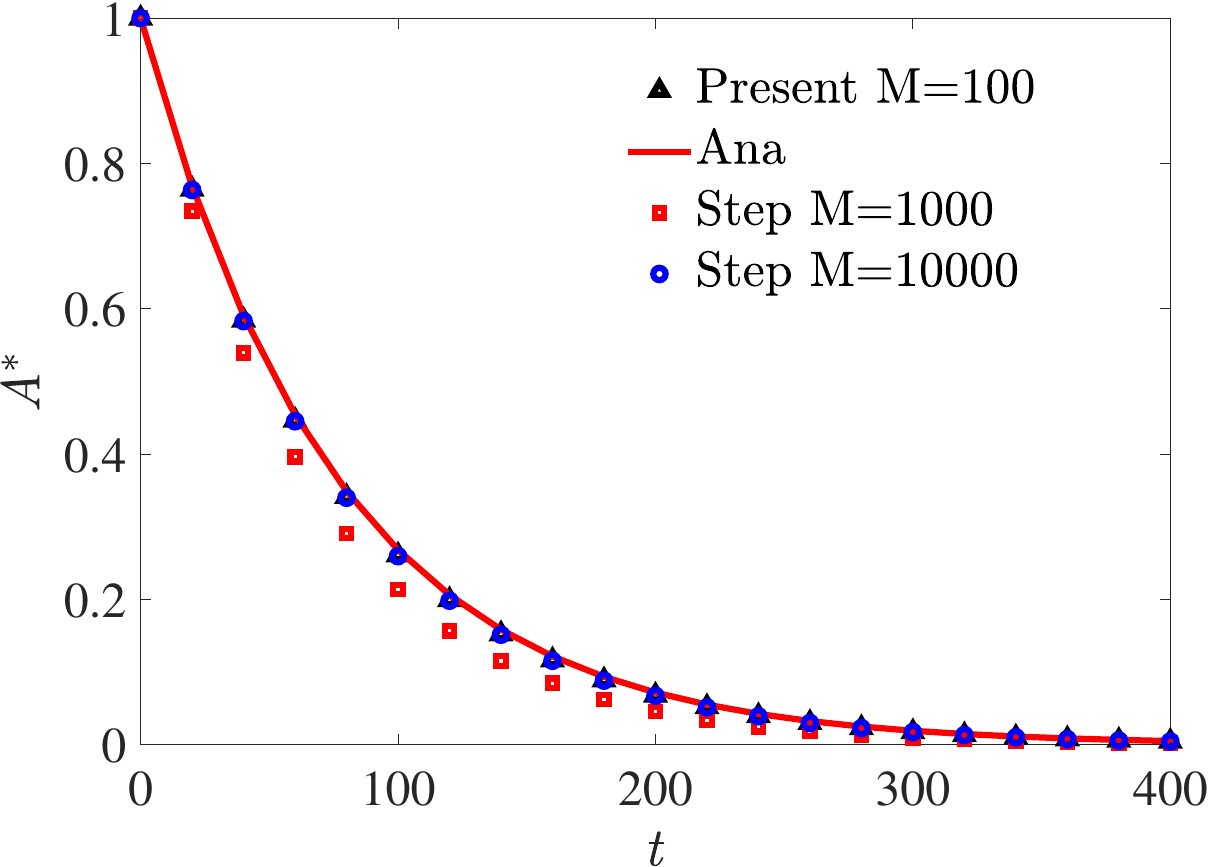}  } ~~
 \subfloat[Kn=0.0001, $\Delta t=0.005, \Delta x=0.01$]{ \label{kn011} \includegraphics[width=0.46\textwidth]{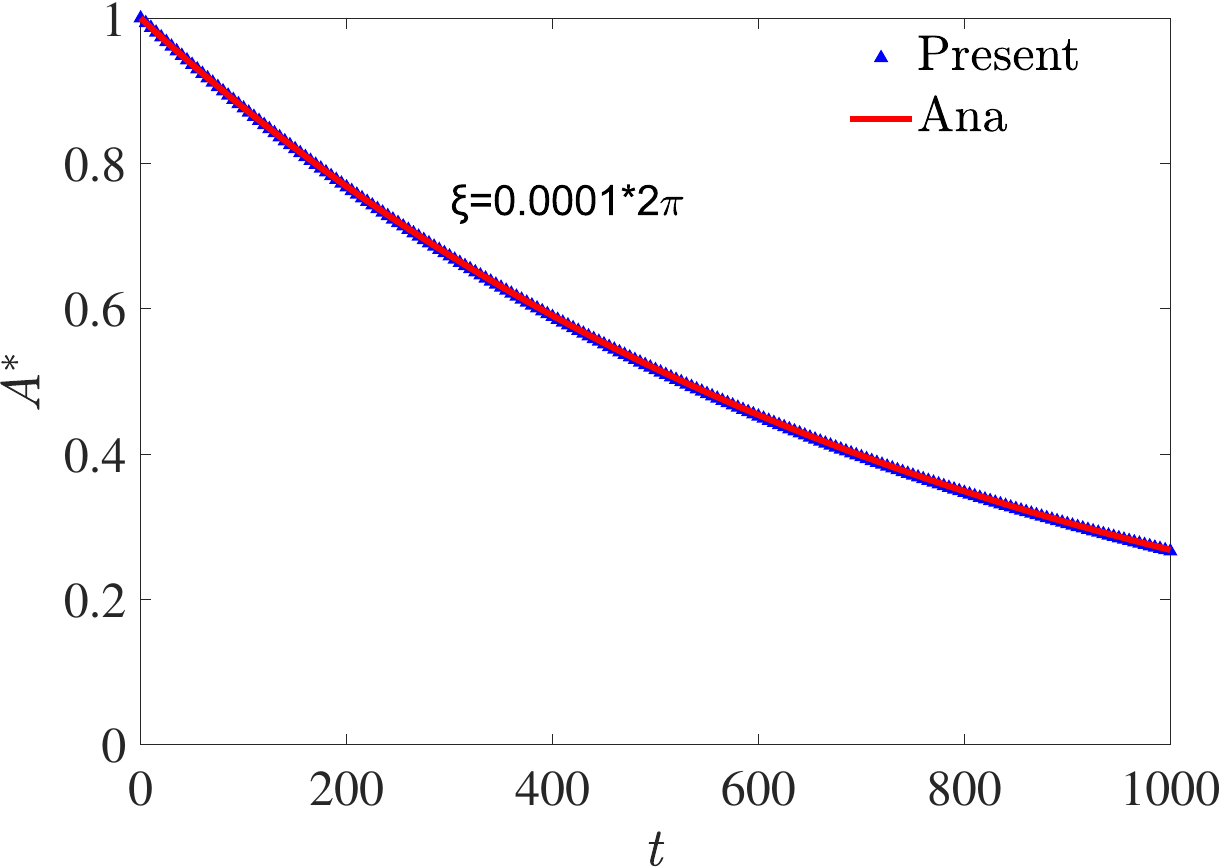}}
 \caption{Decaying of the amplitude of temperature variation with time at different Knudsen numbers, where $A^*=A/A_0$. `Ana' is the analytical solutions~\cite{collins_non-diffusive_2013,GuoZl16DUGKS}. (a) $N_{\theta} =48$. (b,c,d) $N_{\theta} =8$. More discretized parameters can be found in~\cref{lab}. }
 \label{TTGthermal}
\end{figure}
\begin{table}[htb]
\caption{A comparison of time step $\Delta t$, cell size $\Delta x$ and CPU wall time between the present scheme and typical step scheme with forward Euler~\cite{YangRg05BDE,Hu_2024,GuoZl16DUGKS}.}
\centering
\renewcommand\arraystretch{1.5}
\begin{tabular}{ccccccc}
\hline
\hline
      & \multicolumn{3}{c |}{Present}                    & \multicolumn{3}{c}{Step  } \\ \hline
Kn    & $\Delta x$   & $\Delta t$    & \multicolumn{1}{c|}{CPU time (s)} & $\Delta x$       & $\Delta t$       &CPU time (s)  \\ \hline
$10^{-2}$   & 0.01 & 0.005 & \multicolumn{1}{c|}{5.3}    & 0.001    & 0.0005   & 419.2     \\
$10^{-3}$  & 0.01 & 0.005 & \multicolumn{1}{c|}{5.6}    & 0.0001   & 0.00005  & 287702.2  \\ 
$10^{-4}$  & 0.01  & 0.005  & \multicolumn{1}{c|}{46.5}    &  \multicolumn{3}{c}{ unaffordable}   \\ 
\hline
\hline
\end{tabular}
\label{lab}
\end{table}

Numerical simulations are carried out under different Knudsen numbers and they are compared with the analytical solutions derived in Ref.~\cite{collins_non-diffusive_2013},
\begin{equation}
A^* (t^*) = \text{sinc} (\xi t^*) \exp(-t^*) + \int_{0}^{t^*} A^* (t') \text{sinc} \left( \xi (t'-t^*) \right) \exp( t'-t^* )  dt',
\end{equation}
Where $A^*=A/A_0$, $t^*=t/\tau $, $\xi =2\pi$Kn, Kn=$|\bm{v_g}| \tau /L$. 
In cases where the Knudsen number is significantly large ($\xi$=5, 2, 1, 0.5, 0.25), we employ a uniform mesh with $M=50$. 
The results of temperature amplitude variation with time under different Knudsen numbers are shown in~\cref{TTGthermal}, from which it can be found that the present scheme can accurately capture the transient heat conduction from the ballistic regime to diffusive limit.
In addition, a comparison of computational efficiency and accuracy is made in~\cref{lab} and~\cref{TTGthermal} between the present scheme and typical step scheme with forward Euler~\cite{YangRg05BDE,Hu_2024,GuoZl16DUGKS} when the Knudsen number is small, which has only first-order temporal and spatial accuracy.
Numerical results show that the step scheme requires a much finer cell size and smaller time step to get accurate simulation results. 
Its time step has to be smaller than the relaxation time when the Knudsen number is small, otherwise the program will explode.
Hence, it is very hard for the step scheme to accurately simulate the transient heat conduction when Kn$=10^{-4}$ due to unaffordable computational CPU time.
Oppositely, it can be found that even when the time step is much larger than the relaxation time ($\Delta t/\tau=100$), the simulated data predicted by the present scheme in the diffusive regime with Kn$=10^{-4}$ is still accurate, as shown in~\cref{kn011}.
These results show that the semi-implicit Lax-Wendroff kinetic scheme has great advantages than the typical step scheme in terms of computational efficiency and multi-scale heat conduction, especially when the Knudsen number is much small.

\subsection{Quasi-2D in-plane heat conduction} 

Quasi-2D in-plane heat conduction models are usually adopted to characterize the thermal properties of nano-materials, facilitate the optimization of thermal design and ensure that devices work within a safe temperature range.
In the following, we perform a quasi-2D in-plane heat conduction simulation. 
A constant horizontal temperature gradient $dT/dx$ is applied in the in-plane direction, as shown in~\cref{2Dinplane}. 
The distance between the top and bottom boundaries is $H$, and the distance between the left and right boundaries is $L$.
The initial temperature inside the domain is $T_c$.
The left and right boundary temperature is $T_c + dT/dx \times L/2 $ and $T_c -dT/dx \times L/2$, respectively. 
Diffusely reflecting adiabatic boundary conditions are implemented for both the top and bottom boundaries, and periodic boundary conditions are employed for the left and right sides.
\begin{figure}[htb]
     \centering
     \subfloat[]{\includegraphics[width=0.45\textwidth]{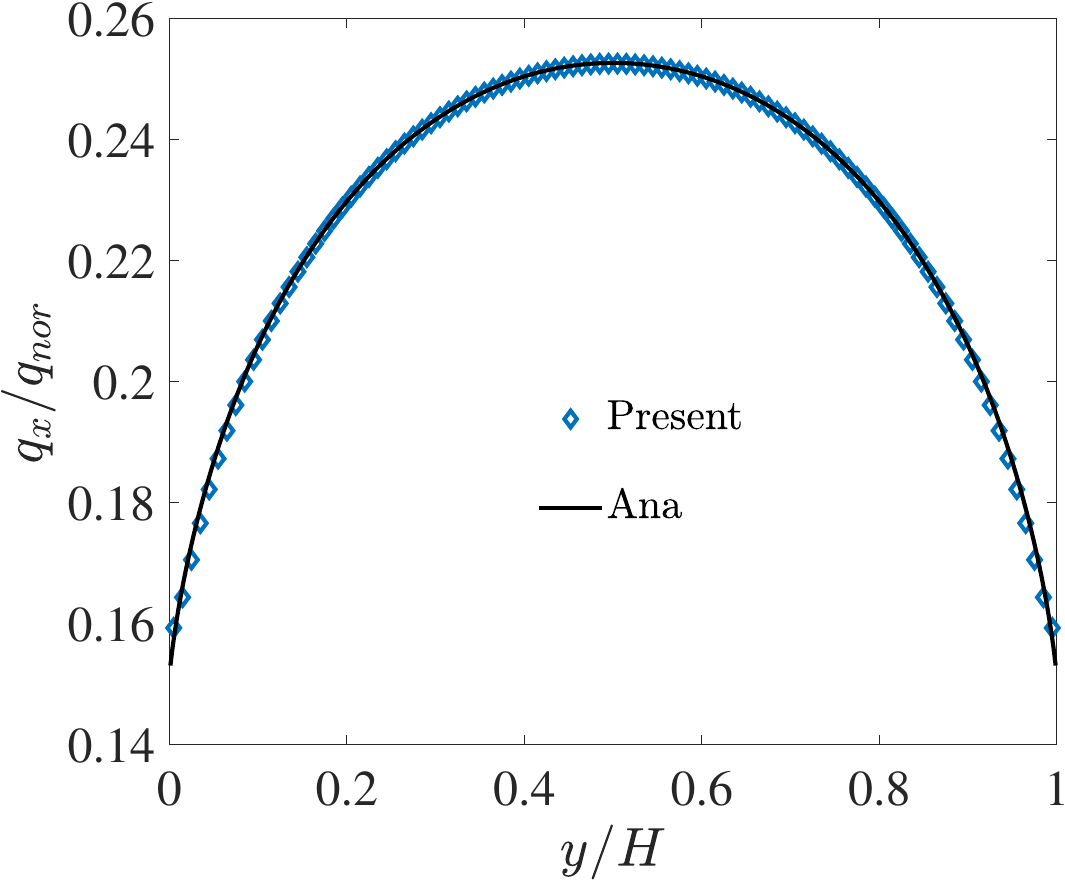}}~~
     \subfloat[]{\includegraphics[width=0.45\textwidth]{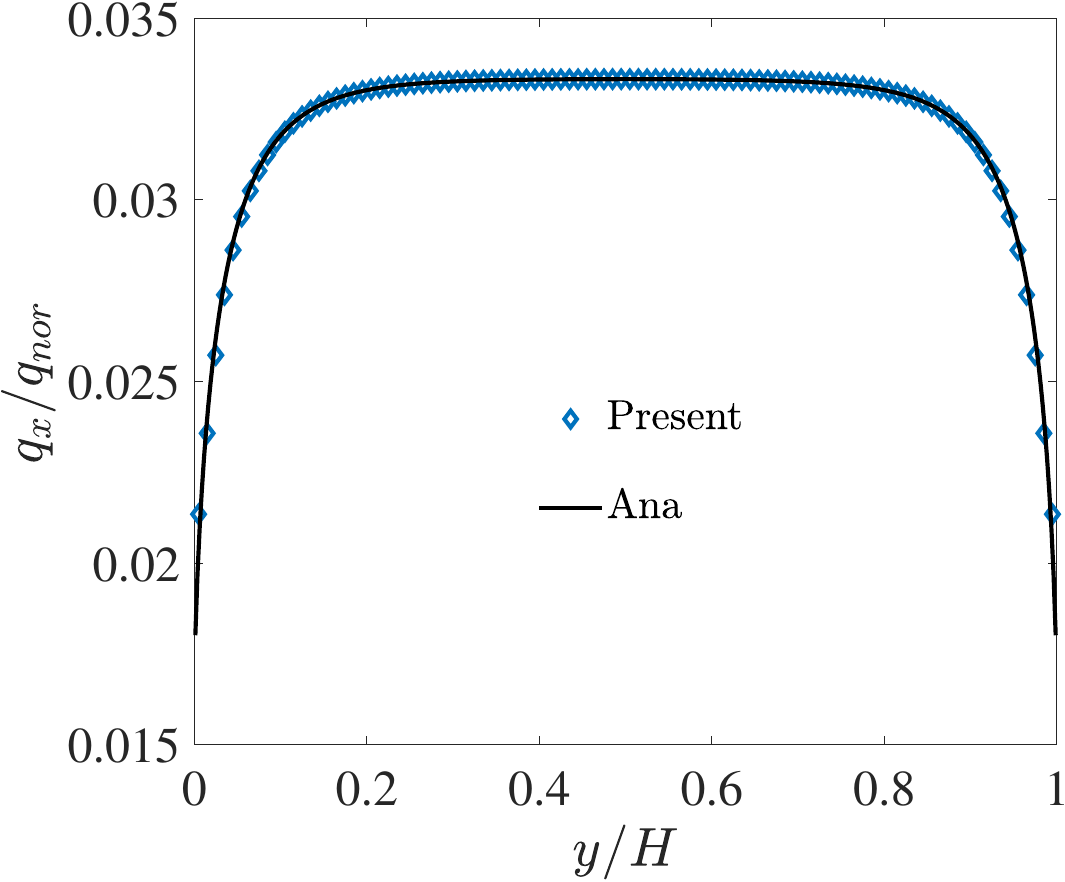}}~~\\
     \subfloat[]{\includegraphics[width=0.45\textwidth]{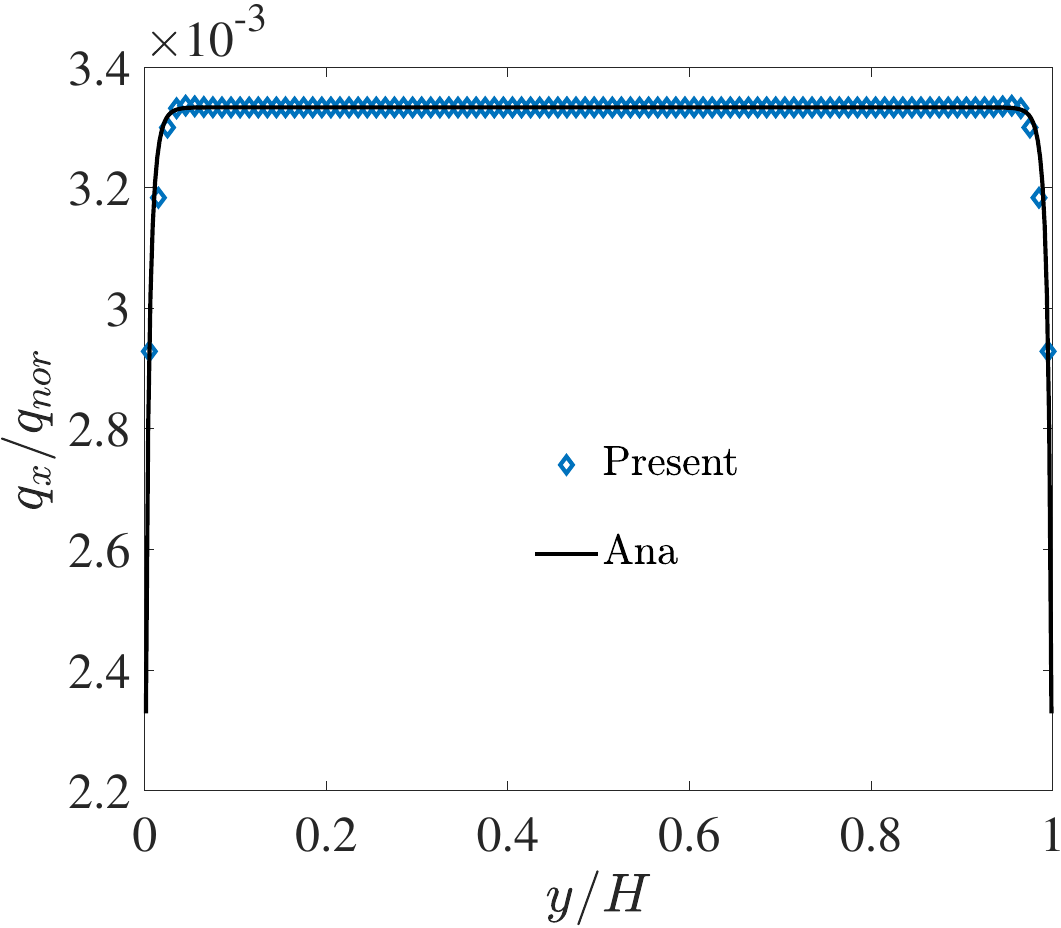}}~~
     \caption{The spatial distributions of the heat flux, where $q_{nor}=-C|{\bm{v_g}}|^ 2dT/dx$, $N \times M = 100 \times 5$, $N_{\theta} \times N_{\varphi} = 48 \times 48$ or $24 \times 24$, `Ana' is the Fuchs-Sondheimer analytical solutions~\cite{Fuchs_1938_analytical,Sondheimer_1952_analytical,turney2010plane}. (a) Kn=1.0, (b) Kn=0.1, (c) Kn=0.01}
     \label{h2d}
\end{figure}

$N \times M = 100 \times 5$ uniform grids are used in the calculation domain.
The system reaches steady state when $\epsilon _2<1.0\times 10^{-10}$, where
\begin{align}
\epsilon _2= \frac{1}{N}\displaystyle\sum_{j=1}^{N}\left | \frac{{q}_{j}^{n+1}-{q}_{j}^{n}}{{q}_{j}^{n}}\right |
\end{align}
The spatial distributions of the heat flux at Kn=1, Kn=0.1 and Kn=0.01 are shown in~\cref{h2d}, respectively.  It can be seen that the numerical results are in good agreement with the Fuchs-Sondheimer analytical solutions~\cite{Fuchs_1938_analytical,Sondheimer_1952_analytical,turney2010plane} 
\begin{align}
q_x\left( Y\right)=-\frac{dT/dx}{4}\int\limits_{0}^{1}C \tau {\lvert u\rvert}^{2}(1-{\eta }^{2})\times \left\{ 2-\exp\left(-\frac{Y}{\eta \text{Kn}}\right)-\exp\left(-\frac{1-Y}{\eta \text{Kn}}\right)\right\}d\eta 
\end{align}

\subsection{Quasi-2D heat conduction geometry}

\begin{figure}[htb]
     \centering
     \subfloat[]{\label{cl1}\includegraphics[width=0.47\textwidth]{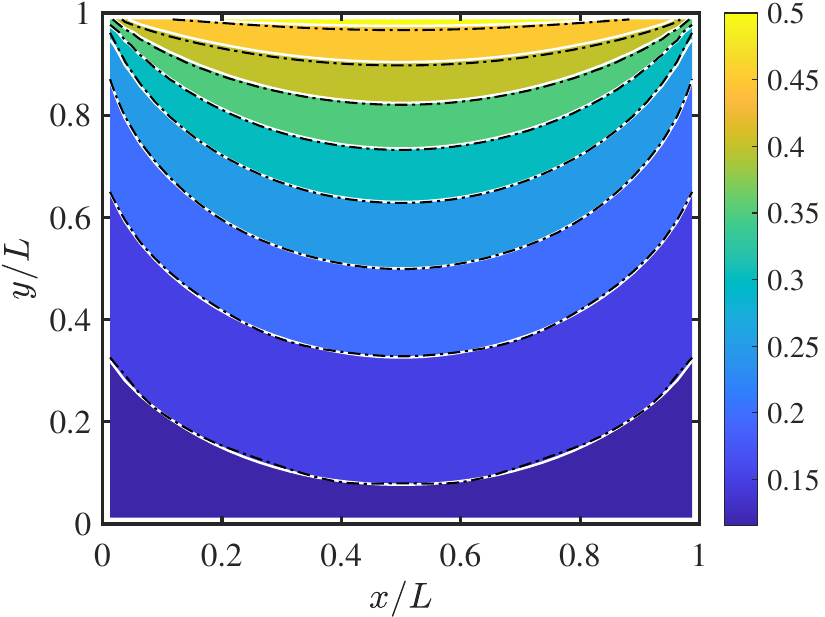}}~~
     \subfloat[]{\label{cl2}\includegraphics[width=0.47\textwidth]{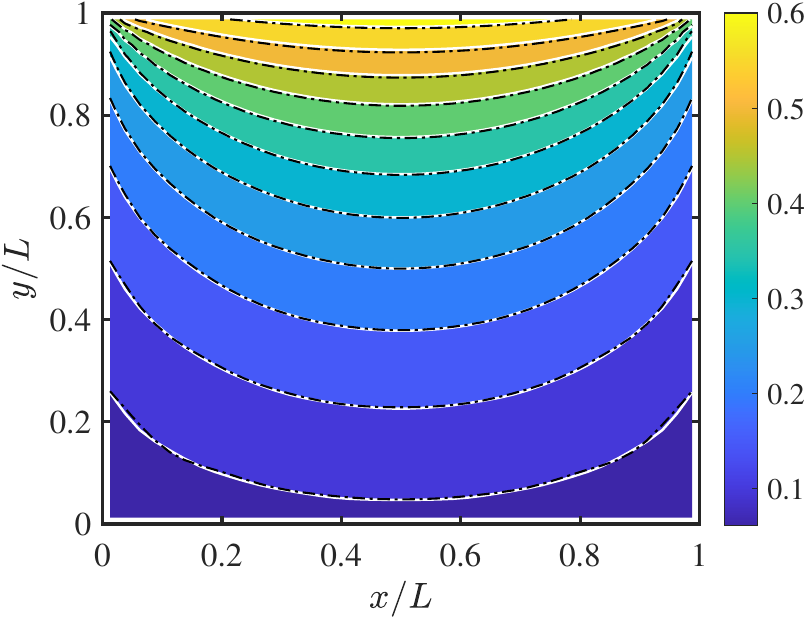}} \\
     \subfloat[]{\label{cl3}\includegraphics[width=0.47\textwidth]{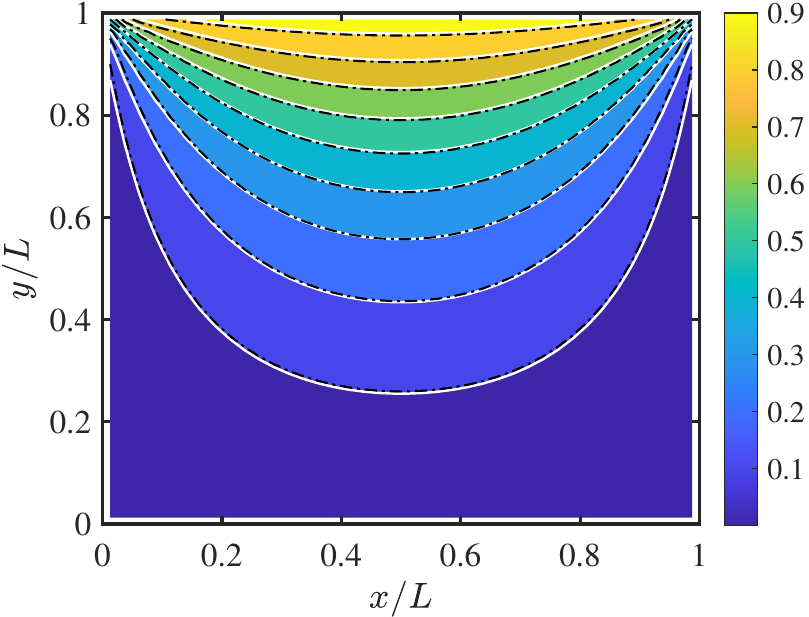}}~~~~
     \subfloat[]{\label{cen}\includegraphics[width=0.47\textwidth]{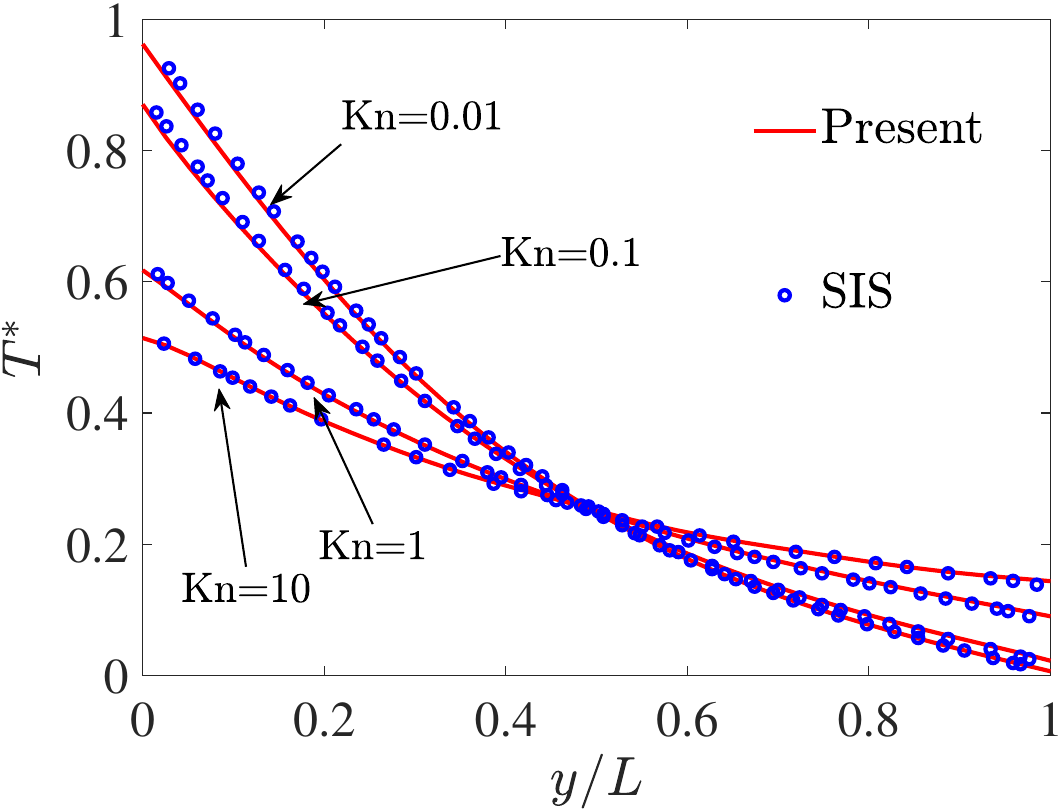}}
     \caption{ (a,b,c) Temperature contour under different Knusdsen numbers Kn=10, Kn=0.1, and Kn=0.01, respectively. When Kn=10, we set $N_{\theta} = N_{\varphi} =48$, $\Delta t=0.008$. When Kn $\leq $1, we set $N_{\theta} = N_{\varphi} =24$, $\Delta t=0.01$. The colored background and white line is the present results, and the dashed line is predicted by the synthetic iterative scheme (SIS)~\cite{ZHANG2023124715}.  (d) The temperature distributions along the vertical central line under different Kn, where $T^*=(T-T_c)/(T_h-T_c)$.}
     \label{2d}
\end{figure}

Quasi-2D heat conduction problem at different wall temperatures is 
ubiquitous in the heat dissipation process of microelectronics~\cite{TCAD_application_intel_2021_review,HUA2023chapter}, and its thermal simulations can offer theoretical guidance for the practical thermal management in chips.
In this paper, the heat conduction in a square geometry with side length $L=1$ is simulated, as shown in~\cref{2Dsquare}. The temperatures of top boundary and the other three boundaries are $T_h$ and $T_c$, respectively, where $T_h>T_c$.
The initial temperature inside the whole computational domain is $T_c$.
Isothermal boundary conditions are applied to all four boundaries.

$N \times M = 40 \times 40$ uniform grids are used in the calculation domain. 
When Kn=10, we used $N_{\theta} \times N_{\varphi} = 48 \times 48$ and when Kn $\leq 1$, we set $N_{\theta} \times N_{\varphi} = 24 \times 24$.
The system reaches steady state when $\epsilon _3<1.0\times 10^{-9}$, where
\begin{align}
\epsilon _3= \frac{1}{N \times M} \displaystyle \sum_{i=1}^{M} \sum_{j=1}^{N}  \left | \frac{{T}_{i,j}^{n+1}-{T}_{i,j}^{n}}{{T}_{i,j}^{n}}\right |.
\end{align}
Based on the aforementioned input parameters, we can obtain the temperature contour in the computational domain, as illustrated in~\cref{cl1,cl2,cl3}.
In addition, the temperature distributions along the $y$ direction in the vertical central line at different Knudsen number are also depicted in~\cref{cen}.
It can be found that the present results are in excellent agreement with the data obtained by the synthetic iterative scheme as reported in a previous paper~\cite{Chuang17gray}.

\section{Conclusion}

A semi-implicit Lax-Wendroff kinetic scheme is developed for numerically solving the transient frequency-independent phonon BTE for all Knudsen numbers, where the phonon scattering and mitigation are coupled together in a single time step.
Different from the characteristic solution or formal integral solution of BTE adopted in the DUGKS or UGKS, the phonon BTE at the cell interface is discretized and solved directly by a finite difference method.
In order to improve both the temporal and spatial accuracies, the trapezoidal and midpoint rules are used to deal with the temporal integration of phonon scattering and convection terms, and second-order upwind and central scheme are used to deal with the spatial interpolation and gradient of interfacial distribution function.
Numerical tests show that the present scheme could accurately predict the steady/unsteady heat conduction in solid materials from the ballistic to diffusive regime, and its time step or cell size is not limited by the relaxation time or phonon mean free path.

\section*{Acknowledgments}

The authors acknowledge Beijng PARATERA Tech CO.,Ltd. for providing HPC resources that have contributed to the research results reported within this paper.
C. Z. acknowledges the members of online WeChat Group: Device Simulation Happy Exchange Group, for the communications on phonon BTE simulations.

\bibliographystyle{elsarticle-num-names_clear}
\bibliography{phonon}
\end{document}